\pdfoutput=1
\documentclass[12pt,a4paper]{article}
\usepackage{xcolor}
\usepackage{enumerate}
\usepackage{framed}
\usepackage{mdframed}
\usepackage{amsmath}
\usepackage{amsfonts}
\usepackage{mathrsfs}
\usepackage{amssymb}
\usepackage{graphicx, rotating}
\usepackage{epsfig}
\usepackage{latexsym}
\usepackage{graphicx}
\usepackage{color}
\usepackage{amsmath,bm,amssymb}
\usepackage{cite}
\usepackage{slashed}
\usepackage{hyperref}
\usepackage{epstopdf}
\usepackage[font=footnotesize,labelfont=]{caption}
\AppendGraphicsExtensions{.tif}
\hypersetup{colorlinks=true, citecolor=bluscuro, linkcolor=black, urlcolor=bluscuro}
\definecolor{rossos}{cmyk}{0,1,1,0.55}
\definecolor{bluscuro}{rgb}{0.15, 0.2, .85}
\definecolor{bluchiaro}{cmyk}{1,.3,0.,0.1}

\def\0{\vec{0}}

\newcommand{\dd}{{\rm d}}

\def\beq{\begin{equation}}
\def\eeq{\end{equation}}

\newcommand{\be}{\begin{equation}}
\newcommand{\ee}{\end{equation}}
\newcommand{\bea}{\begin{eqnarray}}
\newcommand{\eea}{\end{eqnarray}}
\def\beqa{\begin{eqnarray}}

\def\eeqa{\end{eqnarray}}
\def\lsim{\mathrel{\rlap{\lower4pt\hbox{\hskip0.5pt$\sim$}}
    \raise1pt\hbox{$<$}}}         
\def\gsim{\mathrel{\rlap{\lower4pt\hbox{\hskip0.5pt$\sim$}}
    \raise1pt\hbox{$>$}}}         

\numberwithin{equation}{section}

\def\bq{\begin{quote}}
\def\eq{\end{quote}}

 at 10truept

\newcommand{\arXiv}[2]{\href{http://arxiv.org/pdf/#1}{{\tt [#2/#1]}}}
\newcommand{\arXivold}[1]{\href{http://arxiv.org/pdf/#1}{{\tt [#1]}}}

\usepackage[normalem]{ulem}

\begin{document}
\def\thefootnote{\fnsymbol{footnote}}

\begin{center}
\Large{\textbf{Primordial Black Holes from Inflation and Quantum Diffusion}} \\[0.5cm]
\end{center}
\vspace{0.5cm}

\begin{center}

\large{M. Biagetti$^{\rm a}$\footnote{m.biagetti@uva.nl}, G. Franciolini$^{\rm b}$\footnote{gabriele.franciolini@unige.ch}, A. Kehagias$^{\rm c}$\footnote{kehagias@central.ntua.gr} and  A.~Riotto$^{\rm b}$\footnote{antonio.riotto@unige.ch}}
\\[0.5cm]

\small{
\textit{$^{\rm a}$
Institute of Physics, Universiteit van Amsterdam, Science Park, 1098XH Amsterdam, The Netherlands}}

\small{
\textit{$^{\rm b}$Department of Theoretical Physics and Center for Astroparticle Physics (CAP) \\
24 quai E. Ansermet, CH-1211 Geneva 4, Switzerland}}

\small{
\textit{$^{\rm c}$Physics Division, National Technical University of Athens, 15780 Zografou Campus, Athens, Greece}}

\vspace{.2cm}

\end{center}

\vspace{.7cm}

\hrule \vspace{0.3cm}
\noindent \small{\textbf{Abstract}\\ 
Primordial black holes as dark matter may be generated in single-field models of inflation thanks to the enhancement at small scales  of the comoving curvature perturbation.
This mechanism requires leaving the slow-roll phase  to enter  a non-attractor phase during which the inflaton travels across a plateau  and its  velocity drops down 
exponentially. We argue that quantum diffusion has a significant impact  on the  primordial black hole mass fraction making the classical standard prediction not trustable.
}

\vspace{0.3cm}
\noindent
\hrule
\def\thefootnote{\arabic{footnote}}
\setcounter{footnote}{0}

\renewcommand{\theequation}{1.\arabic{equation}}
\setcounter{equation}{0}
\vspace{0.9cm}
\section*{I. Introduction}
\noindent
The interest in the physics of Primordial Black Holes (PBHs) and the possibility that they form all (or a fraction) of the dark matter in the universe has risen up  \cite{PBH1,PBH2,PBH3,kam,rep1,rep2,revPBH,c0,c1,c2}  again after the discovery of two $\sim 30 M_\odot$ black holes through the gravitational  waves generated during their merging \cite{ligo}. A  standard  mechanism to account for the generation of the  PBHs is  through the boost  of the curvature perturbation ${\cal R}$   at small scales   \cite{s1,s2,s3}. Such an enhancement  can occur either within single-field models of inflation, see Refs  \cite{sone1,sone0,sone2,sone3,sone4} for some recent literature,   or through some spectator field  \cite{stwo1,Carr,gauge,cs} which could be identified   the Higgs of the Standard Model \cite{ssm2}.

In order for this primordial mechanism to occur, one needs an enhancement of the  power spectrum of the curvature perturbation  from its $\sim 10^{-9}$ value at large scales to $\sim 10^{-2}$ on small scales. Subsequently, these  large perturbations are   communicated to radiation during the reheating process after inflation and they may give rise to PBHs upon horizon re-entry if they are  sizeable enough. If we indicate by ${\cal P}_{\cal R}$
 the comoving curvature  power spectrum, 
a  region of size the Hubble radius may collapse and form a  PBH if   the corresponding square root of the variance $\sigma_{\cal R}$ smoothed with a  high-pass filter on the same length scale (comoving momenta larger than the inverse of the comoving Hubble)   is larger than 
some critical value ${\cal R}_c$.  Its  exact value is sensitive  to the  equation of state upon horizon re-entry and it is about 0.086 for radiation \cite{harada}. However, 
 larger values have been adopted  in the literature  \cite{bb,miller,hu}. We will later on use the common representative value   ${\cal R}_c\simeq 1.3$.

Under the (strong) hypothesis  that the curvature perturbation   obeys a Gaussian statistics, the  primordial mass fraction $\beta_{\rm prim}(M)$ of the universe occupied by  PBHs formed at the time  of formation   reads
\be
\beta_{\rm prim} (M)=\int^\infty_{{\cal R}_c} \frac{{\rm d}{\cal R}}{\sqrt{2\pi}\,\sigma_{\cal R}}e^{-{\cal R}^2/2\sigma_{{\cal R}}^2}.
\ee
It  corresponds to a present
dark matter abundance made of PBH of masses $M$ given by (neglecting accretion) \cite{hu}
\be
\left(\frac{\Omega_{\rm DM}(M)h^2}{0.12}\right)\simeq\left(\frac{\beta_{\rm prim}(M)}{7\cdot 10^{-9}}\right)\left(\frac{\gamma}{0.2}\right)^{1/2}\left(\frac{106.75}{g_*}\right)^{1/4}\left(\frac{M_\odot}{M}\right)^{1/2},
\ee
where $\gamma<1$ is a parameter accounting for the efficiency of the collapse and  $g_*$ is the effective number of degrees of freedom. Imposing PBHs to be  the dark matter,  values $\gamma<1$ require larger values of 
$\beta_{\rm prim}$ and therefore to be  conservative we impose $\gamma=1$ \cite{hu}
\be
\beta_{\rm prim}(M)\gsim 3\cdot 10^{-9}\,\left(\frac{M}{M_\odot}\right)^{1/2}.
\ee
For a mass of the order of $10^{-15}\, M_\odot$ we find   \cite{hu}
\be
\label{small}
\sigma_{{\cal R}}\gsim 0.16.
\ee
 Now, in   single-field models of inflation the power spectrum of the comoving curvature perturbation is given by (we set the Planckian mass equal to one from now on) \cite{lrreview}
\be
\label{gh}
{\cal P}^{1/2}_{\cal R}(k)=\left(\frac{H}{2\pi\phi'}\right), \,\,\,\,\phi'= \frac{\dd\phi}{\dd N},
\ee
where $N$ is the number of e-folds,  the prime denotes differentiation with respect to $N$, and  $H$ is the Hubble rate. The generation of PBHs requires the jumping within a few e-folds $\Delta N$ of the value of the power spectrum of about seven orders of magnitude from its value on CMB scales. Without even specifying the single-field model of inflation,  one may conclude that there must be a  violation of the slow-roll condition as $\phi'$ must change rapidly with time. This may happen  when the inflaton field goes through  a so-called non-attractor phase  (dubbed also ultra-slow-roll) \cite{u1,u2,u3,u3.1,u4,u4.1,u5,u6,pajer,sasaki} in the scalar potential, thus   producing a sizeable resonance in the power spectrum of the curvature perturbation.

When the inflaton experiences a plateau in its potential, since $\phi'$ must be extremely small, a short non-attractor period  is achieved during which  the equation of  motion
of the inflaton background $\phi$ reduces to
\be
\phi''+3 \phi'+\frac{V_{,\phi}}{H^2}\simeq \phi''+3 \phi'= 0,
\ee
where ${}_{,\phi}$  denotes differentiation with respect to the inflaton field $\phi$ with potential $V(\phi)$. The comoving curvature perturbation  increases, due to  its decaying mode which in fact is  growing,  as 
\be
\phi'\sim e^{-3N}\,\,\,{\rm and}\,\,\, {\cal P}^{1/2}_{{\cal R}} \sim e^{3N}.
\ee
It is this exponential growth which helps  obtaining large fluctuations in the curvature perturbation and the formation of PBHs upon horizon re-entry during the radiation phase.
This is also the reason why the power spectrum should be  
quoted at the end of inflation and not, as usually done in slow-roll, at Hubble crossing:  its  value at the instant of Hubble  crossing  differs by a significant factor from the asymptotic value at late times. Not respecting these rules might lead to an incorrect estimate of the power spectrum and consequently of the PBH abundance at formation, see e.g.  Ref. \cite{ssm1} and the subsequent discussions in Refs. \cite{u5,hu}.

Putting aside the strong sensitivity of the PBH mass fraction at formation  to possible non-Gaussianities \cite{s3,byrnes,ngtwo,ng1,ng2,ng3,ng33,ng4,ng5,ng6,ng7} which is common to all mechanisms giving rise to PBHs through sizeable perturbations (we will however devote Appendix C for some considerations about non-Gaussianity where
we will  show that the $\delta N$ formalism \cite{deltan} can help in assessing the
role of non-Gaussianity for those perturbations generated during the non-attractor phase) in this paper we are interested in another issue, the  role
of quantum diffusion. 
 One might be reasonably suspicious  that  during the non-attractor phase the quantum diffusion becomes relevant \cite{ng6}. 
The reason is the following. 
%
%
%
%
%
The stochastic equation of motion for the classical inflaton  field takes into account that each Hubble time the inflaton field receives kicks of the order of $\pm(H/2\pi)$ \cite{linde}
\be
\label{stoch}
\phi''+3 \phi'+\frac{V_{,\phi}}{H^2}=\xi,
\ee
where $\xi$ is  a Gaussian random noise with
\be
\langle\xi(N)\xi(N')\rangle=\frac{ 9 H^2}{4\pi^2}\delta(N-N').
\ee
During the non-attractor phase, $V_{,\phi}$ needs to be tiny enough to allow $\phi'$ to promptly decrease thus  violating slow-roll. For the same reason, one needs to make sure that quantum jumps are not significant in this case. One could try to impose the condition
\be
\label{mk}
\frac{2\pi V_{,\phi}}{3H^3}\gsim  1.
\ee
to be satisfied during the non-attractor phase. In slow-roll, the condition (\ref{mk}) could be exactly expressed in terms of the power spectrum and the latter would be required to be smaller than unity, thereby giving a direct constraint on a physical observable. However, during the non-attractor phase, the
bound (\ref{mk}) is not directly expressed in terms of the power spectrum (\ref{gh}), making the comparison with physical observables more difficult. One might naively think that
during the non-attractor phase the noise is not relevant if $\phi'$ is larger than $(H/2\pi)$ \cite{sone2}. However, this is not correct for two reasons. First because, as we will see, the relevant noise to be evaluated is the one  for the inflaton velocity. Secondly,  and above all, because this is not the right criterion to evaluate the  strong impact of quantum diffusion   onto  the PBH mass fraction. 



We will first elaborate on the computation of the power spectrum during the non-attractor phase in order to understand
some basic features of the power spectrum itself, e.g. its time evolution and the location of its peak. This will be useful for the considerations about the quantum diffusion. We will then use 
the so-called Kramers-Moyal equation \cite{Risken} to assess the impact of quantum diffusion. 

The Kramers-Moyal equation is the suitable starting point as it
highlights the importance of the inflaton velocity and it is a generalisation of the Fokker-Planck equation. Indeed, 
in general the Kramers-Moyal equation contains   an infinite number of derivatives with respect to the inflaton field after having integrated out the velocity,  while the Fokker-Planck equation is a truncation of the Kramers-Moyal equation by retaining only    two spatial derivatives of the inflaton field. This is not an irrelevant point: 
 Pawula's theorem \cite{pawula}  tells us that if we set to zero some  coefficient  $c_n$ of the higher derivative terms with respect to the inflaton field with $n\geq 3$, then  all the coefficients of the higher derivatives are   zero. It is therefore not  consistent to keep some higher derivative term unless 
all of them  are kept. This means that, whenever the Kramers-Moyal equation may not be solved exactly, a numerical approach is useful to solve it without applying  an unjustified truncation.

We will present both analytical and numerical results to quantify the impact of quantum diffusion on the PBH abundance. In the simplest case of non-attractor phase with an approximately   linear  or quadratic potential, the system can be solved analytically. For a linear potential, the stochastic motion of the system is  characterised by the fact that the  variance of the  velocity  $\phi'$ rapidly converges towards  a stationary value. For a quadratic potential, the spread in the velocity varies with time,  although slowly,  away from the stationary point.

However,  for more complicated situations, e.g. if  during the non-attractor phase the inflaton goes through  an  inflection point, a numerical analysis is called for and we will show that  the spread in the velocity grows
with time. If this growth is too large, classicality as well as information on the PBH abundance is lost. 
We will propose  two criteria to be respected in order to neglect the quantum diffusion and we will  
see that the curvature perturbation is severely constrained from above
in order to avoid an undesirable spreading of the velocity wave packet. We will also argue  that the capability of the standard (that is classical) calculation to predict the 
 correct dark matter abundance  in terms of PBH is severely challenged by the presence of quantum diffusion.


 This paper is organised as follows. In section II we
 discuss the computation of the curvature perturbation during the non-attractor phase. We then  start our study of the quantum diffusion, both analytically and numerically,  in section III and IV respectively. 
 In sections V and VI we offer two criteria to assess the importance of the diffusion. Finally, in section VII we offer our conclusions.
 
 The paper contains several Appendices. Appendix A deals with the curvature perturbation, the Schwarzian derivative and the dual transformation; appendix B with the study of the
 evolution of the comoving curvature perturbation from the non-attractor phase back to the slow-roll phase; appendix C offers some consideration about non-Gaussianity.
\renewcommand{\theequation}{2.\arabic{equation}}
\setcounter{equation}{0}
\section*{II. The   comoving curvature perturbation and the non-attractor phase}\label{secII}
\noindent
In this section we offer some considerations about the curvature perturbation generated thanks to the non-attractor phase. We start by some analytical considerations and then we will proceed with a more realistic example. 

\subsection*{Non-attractor: some analytical considerations}
\noindent
We are interested in the curvature perturbation for those modes  leaving the Hubble radius deep in the non-attractor phase. We suppose that the non-attractor
phase starts when the inflaton field acquires the value $\phi_0$ and ends when it becomes equal to $\phi_\star$. We also assume that the non-attractor
phase is preceded and followed by slow-roll phases, see Fig. 1. During these phases $\eta=-\phi''/\phi'$ passes from a tiny value (slow-roll) to 3 (non-attractor) back to small values
(slow-roll).
\noindent
\begin{figure}[ht!]
    \begin{center}
      \includegraphics[scale=1]{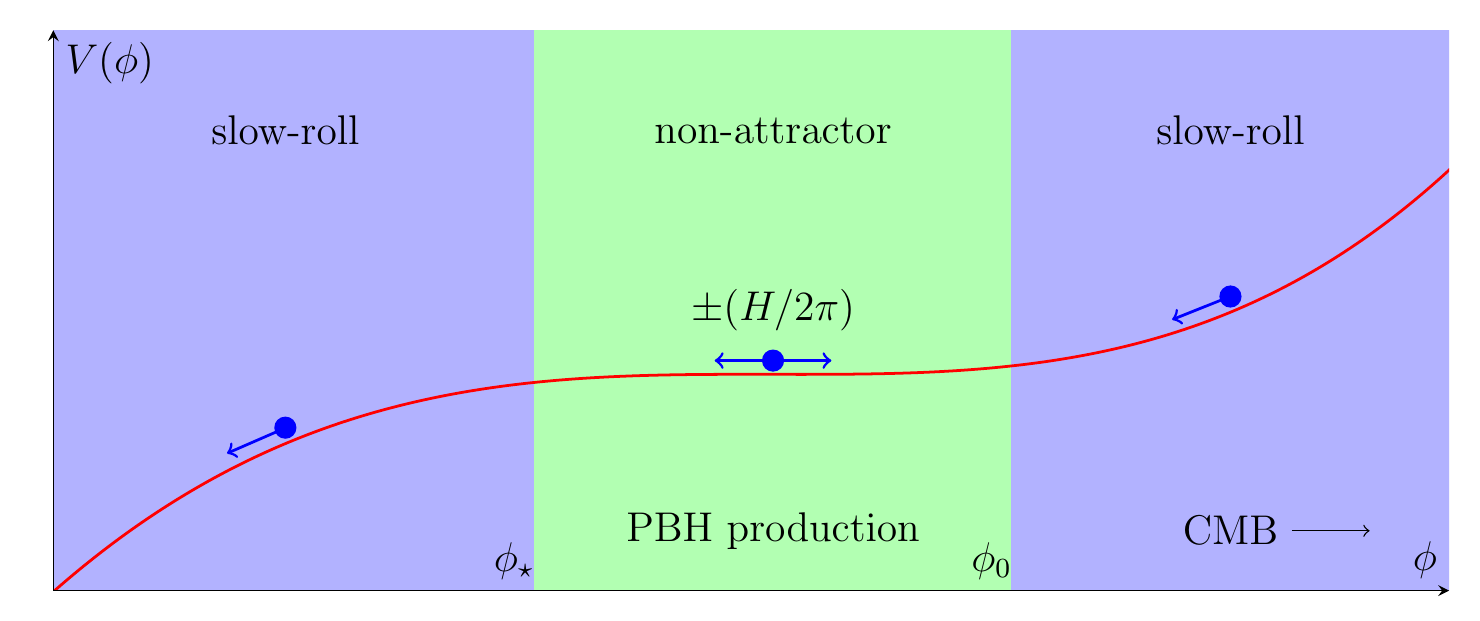}
    \end{center}
    \label{one}
     \caption{\small A representative behaviour of the inflaton potential during the various phases, highlighting the possibility of quantum diffusion during the non-attractor phase.}  
\end{figure}
To treat the problem analytically, we suppose that during the non-attractor phase  we may Taylor expand the inflaton potential as 
\be
\label{linear}
V(\phi)\simeq V_0\left(1+\sqrt{2\epsilon_V}(\phi-\phi_0)\right)+\cdots\,\,\,\,{\rm for}\,\,\,\,\phi_\star<\phi<\phi_0,
\ee
where $\sqrt{2\epsilon_V}=V_{,\phi}/V_0$ is the slow-roll parameter.  Of course, the potential during the non-attractor phase may be more complex, but  its linearisation captures the main features.

Computing everything in terms of the number of e-folds $N$
and defining $N=0$ the beginning of the non-attractor phase with initial conditions $\phi_0$ and $\dd\phi/\dd N|_0=\Pi_0$ and setting $\Pi(N)=\phi'(N)$, the solution of the equations of motion  leads to 
\begin{eqnarray}
\phi(N)&=&\phi_0+\frac{1}{3}(\Pi_0-\Pi)-\sqrt{2\epsilon_V}N,\nonumber\\
\label{eq:pin}
\Pi(N)&=&\sqrt{2\epsilon_V}\left(e^{-3N}-1\right)+\Pi_0 e^{-3N}.
\end{eqnarray}
We see that if the inflaton field starts with a large velocity from the preceeding slow-roll phase, there is a period over which the velocity of the
inflaton field decays exponentially. Depending on the duration of the non-attractor phase, the velocity may or may not attain its slow-roll asymptote
given by $-\sqrt{2\epsilon_V}$. Indicating  by    $\Pi_\star$ the value of the velocity at the end of the non-attractor phase, the final value of the curvature perturbation at the end of the  non-attractor phase
is given by 
\be
\label{po}
{\cal R}_\star=-\left(\frac{\delta\phi}{\Pi}\right)_\star, \,\,\,\,\delta\phi(k)=\frac{H}{\sqrt{2k^3}}.
\ee
The corresponding  power spectrum is flat. This might come as a surprise as slow-roll is badly violated, but in fact its a direct consequence of the dual symmetry described  in Ref. \cite{wands} (see also Refs. \cite{seto,leachetal,ll}). We elaborate extensively  on this point in Appendix A. 
In a nutshell and alternatively, one can show it in the following way. Using the conformal time $\tau$ and setting ${\cal R}=u/z$, $z=(a\,{\rm d}\phi/{\rm d}\tau)/{\cal H}$, where ${\cal H}$ is the Hubble rate in conformal time, one can write the equation for the function $u$ as (cfr. Eq. (\ref{zs}))
\be
\frac{1}{u}\frac{{\rm d}^2u}{\dd \tau^2}=2a^2H^2\left(1+\frac{5}{2}\epsilon+\epsilon^2-2\epsilon\eta-\frac{1}{2}\frac{V_{,\phi\phi}}{H^2}\right)-k^2,
\ee
where $\epsilon$ and $\eta$ are the slow-roll parameters defined in Eq. (\ref{sr}) and $a$ is the scale factor. Since during the non-attractor phase
$\epsilon\ll 1$, $\eta\simeq 3$ and the potential is very flat, the right-hand side of the previous equation is approximated on super-Hubble scales to $2/\tau^2 \simeq ({\rm d}^2a/\dd \tau^2)/a$ and therefore $u\propto a$. It  provides the standard solution for the mode function of the curvature perturbation
\be
{\cal R}_k=\frac{\cal H}{(\dd\phi/\dd\tau)\sqrt{2 k^3}}(1+i k\tau)e^{-ik\tau}.
\ee
This is the standard slow-roll solution with the crucial exception that the inflaton velocity changes rapidly with time.
Notice also that the expression (\ref{po}) can be extended at the end of inflation. This is possible if the transition from the non-attractor phase to the subsequent slow-roll phase (if any) is sudden, i.e. the velocity during the subsequent slow-roll
phase is much bigger than $\Pi_\star$. Under these circumstances, the power spectrum does not have time to change and remains indeed (\ref{po}) till the end of inflation \cite{sasaki}. We give more  details in  Appendix B. 

So far, we  have discussed the  perturbation associated to the modes which leave the Hubble radius deep in the non-attractor phase. However, the peak of the
curvature perturbation is in fact reached for those modes which leave the Hubble radius during the sudden transition from the slow-roll phase into the non-attractor phase. During this
transition the (would-be) slow-roll parameter $\eta=-\Pi'/\Pi$ jumps from a tiny value to 3. 

 To see what happens, we model the parameter $\eta$ as $\eta\simeq 3\theta(\tau-\tau_0)$, where we have now turned again to conformal time $\tau$. If so, and if we indicate
 by $\epsilon_+$ the slow-roll parameter during the slow-roll phase preceding the non-attractor phase and  assume it to be constant in time, one has 
\be
{\cal R}_k=\frac{H}{\sqrt{2\epsilon_+ k^3}}(1+i k\tau)e^{-ik\tau}\,\,\,\,{\rm for}\,\,\,\, \tau<\tau_0,
\ee
and \cite{sasaki}
\be
\left(\frac{\tau}{\tau_0}\right)^3{\cal R}_k=\alpha_k\frac{H}{\sqrt{2 k^3}}(1+i k\tau)e^{-ik\tau}+\beta_k\frac{H}{\sqrt{2 k^3}}(1-i k\tau)e^{ik\tau}\,\,\,\,{\rm for}\,\,\,\, \tau>\tau_0,
\ee
where we have taken into account that immediately after the beginning of the  non-attractor phase the curvature perturbation increases as the inverse cubic power of the conformal time (cfr. Eq. \eqref{eq:pin}).
Imposing continuity of the two functions together of their derivatives, one obtains a power spectrum at the end of inflation
\begin{eqnarray}
\label{peak}
{\cal P}_{\cal R}&=&g(-k\tau_0)\, {\cal P}_{{\cal R}_\star},\nonumber\\
g(x)&=&\frac{1}{2x^6}\left(9 + 18 x^2 + 9 x^4 + 2 x^6 + 3 (-3 + 7 x^4) \cos 2x - 
   6 x (3 + 4 x^2 - x^4) \sin 2x\right).
\end{eqnarray}
The function $g(x)$ is ${\cal O}(x^4)$ for $x\simeq 0$, has a maximum of about  2.5  around $x\simeq 3$ and oscillates rapidly around 1 for $x\gg 1$. 
We can conclude that the  power spectrum has the following shape: it increases, reaches a  peak, and then decreases a bit till  a plateau is encountered. This is in good agreement with what obtained, for instance,  in Refs. \cite{ll,leachetal}.
The amplitude of the peak is about  2.5 times larger than the plateau  in correspondence of  the modes which leave the Hubble radius during the 
non-attractor phase
\be
{\cal P}_{{\cal R}_{\rm pk}}\simeq 2.5\,{\cal P}_ {{\cal R}_\star}=2.5\left(\frac{H}{2\pi\Pi}\right)_\star.
\ee
Of course, given the assumption of sudden transition of $\eta$ from tiny values to $3$ around $\tau_0$ and having assumed $\epsilon_+$ constant, we expect this number to change by a factor ${\cal O}(1)$
depending upon the exact details. The linearisation of the potential is  an approximation, but it captures the main features of the final result. For more complicated situations, for instance if during the
non-attractor phase the inflation crosses an inflection point, one expects again a peak in the curvature perturbation for that mode leaving the Hubble radius at the sudden transition between the
slow-roll and the non-attractor phase. However, one does not expect a significant plateau following the maximum as the non-attractor phase is typically very short. This also implies that the exact amplitude of the peak  depends on the fine details of the transition.

\subsection*{An example: Starobinsky's model.} 
In order to assess the quality of our findings we can consider Starobinsky's model \cite{Starobinsky92}
which is characterised by a potential with two linear regions
\begin{align}
	V(\phi)&\simeq V_0\left(1+\sqrt{2\epsilon_{+}}(\phi-\phi_0)\right)+\cdots\,\,\,\,{\rm for}\,\,\,\,\phi>\phi_0,
\\
V(\phi)&\simeq V_0\left(1+\sqrt{2\epsilon_{-}}(\phi-\phi_0)\right)+\cdots\,\,\,\,{\rm for}\,\,\,\,\phi<\phi_0,
\end{align}
where $\epsilon_{\pm}$ are the slow-roll parameter during the slow-roll phase and the non-attractor phase, respectively.  In fact, to deal with the problem numerically
we have parametrised the discontinuity in the potential as
\be
V(\phi)= V_0\left(1+\frac{1}{2} \left(\sqrt{2\epsilon_{+}}-\sqrt{2\epsilon_{-}} \right)(\phi-\phi_0)\tanh\left(\frac{\phi-\phi_0}{\delta} \right)+ \frac{1}{2} \left(\sqrt{2\epsilon_{+}}+\sqrt{2\epsilon_{-}} \right)(\phi-\phi_0)\right),
\ee
where $\delta$ determines the size of the region in which the potential smoothly changes slope. If $\delta \ll 1$ the potential during the non-attractor phase becomes exactly linear. We will comment on the effect of varying $\delta$ on  the  quantum diffusion in Sec. IV. 

If $\epsilon_{+}\gg \epsilon_{-}$, a  prolonged non-attractor phase is obtained during which 
\be
\frac{\phi'}{{\cal H}}=-\sqrt{2\epsilon_{-}} -(\sqrt{2\epsilon_{+}}-\sqrt{2\epsilon_{-}})(\tau/\tau_0)^3.
\ee
   \begin{figure}[ht!]
 \centering
 \begin{minipage}{0.49\textwidth}
      \includegraphics[scale=0.9]{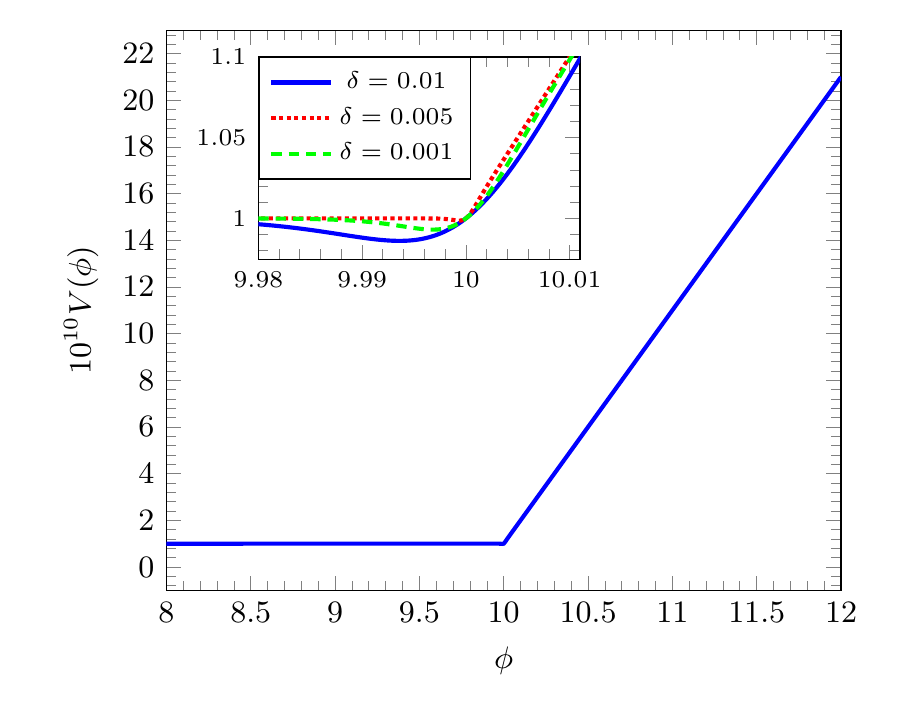}
      \end{minipage}
    \begin{minipage}{0.49\textwidth}
      \includegraphics[scale=0.9]{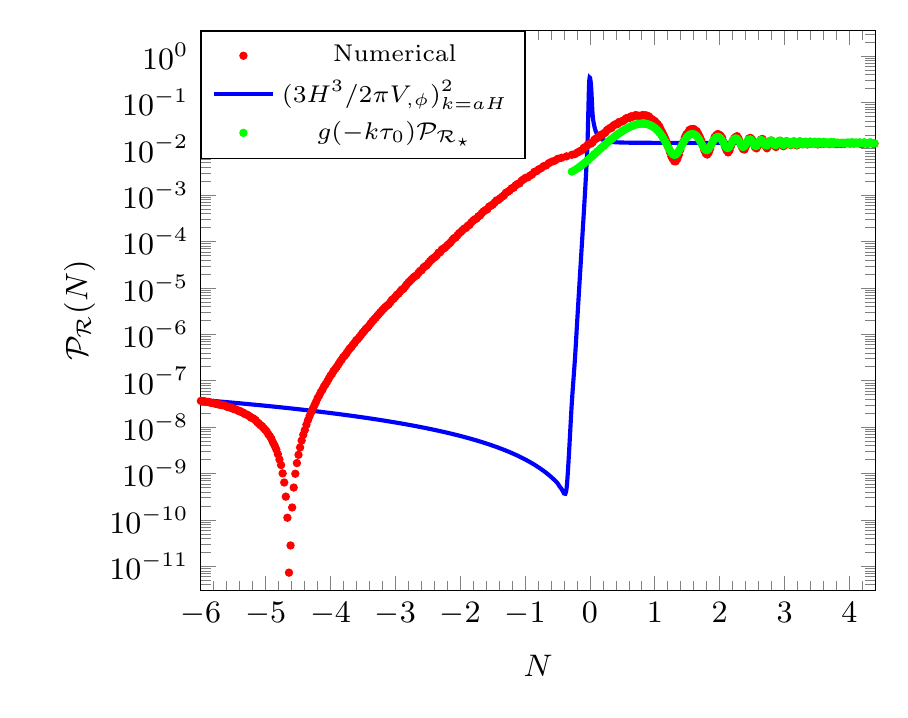}
      \end{minipage}
     \caption{\small On the left, the Starobinsky's potential. On the right, the power spectrum in Starobinsky's model (not normalised at the CMB on large scales) obtained for $\epsilon_+/\epsilon_-=10^8$ and  as a function of $N$ corresponding to $k=aH$. We have arbitrarily set $N=0$ at the time at which $\eta$ reaches 3.}      
     \label{figpostar}
     \end{figure}
\noindent
The inflaton velocity at the beginning of the non-attractor phase   is $-\sqrt{2\epsilon_{+}}$, then it  quickly decays reaching a maximum (recall velocities are negative) at $(\tau_{\rm max}/\tau_0)^3\simeq 2\sqrt{\epsilon_+/\epsilon_-}$ and then it reaches the value $-\sqrt{2\epsilon_{+}}$. The corresponding power spectrum  in Fig. \ref{figpostar} illustrates three relevant points: the fact that the power spectrum reaches a plateau and becomes scale-independent, the amplitude of the plateau is reproduced by the standard slow-roll formula (see Appendix A) and finally that the formula (\ref{peak}) provides a good fit 
to the numerical result.

\subsection*{Non-attractor:  more physical cases}
\noindent
We turn now the discussion to more physical cases discussed recently in the literature.  First, we consider the model in Ref \cite{sone3}. Leaving aside the details of the particular string model giving rise to it, the inflaton potential reads
 \be
 V(\phi)=\frac{{W_0}^2}{{\mathcal{V}}^3} \left[
   \frac{{c_{\rm up}}}{\sqrt[3]{{\mathcal{V}}}}+
   \frac{{a_{\rm w}}}{{e^{\frac{ \phi  }{\sqrt{3}}}}-{b_{\rm w}}}
   -\frac{{c_{\rm w}}}{{e^{\frac{ \phi }{\sqrt{3}}}}}
   +\frac{e^{\frac{2
   \phi }{\sqrt{3}}}}{{{\mathcal{V}}}} \left({d_{\rm w}}-\frac{{g_{\rm w}}}{r_{\rm w}e^{\sqrt{3} \phi}/\mathcal{V} +1}\right)\right],
   \ee
   where the  parameters used in our  analysis can be found in Tab. \ref{tab1}.
   \begingroup
\setlength{\tabcolsep}{10pt} 
\begin{table}[ht!]
\begin{center}
\caption{{\small Parameter set for the string model  in Ref. \cite{sone3}.}}\label{tab1}
\begin{tabular}{|c|c|c|c|c|c|c|c|c|}
\hline
$a_{\rm w}$&$b_{\rm w}$& $c_{\rm w}$&$d_{\rm w}$&$g_{\rm w} $&$r_{\rm w}$&$\mathcal{V}$&$W_0$&$c_{\rm up} $\\
\hline
0.02&1&0.04&0&$3.076278 \cdot 10^{-2}$& $7.071067\cdot 10^{-1}$&$1000$&$12.35$&$0.0382$\\
\hline
\end{tabular}
\end{center}
\end{table}%
\endgroup

\begin{figure}[ht!]
 \centering
    \begin{minipage}{0.49\textwidth}
      \includegraphics[scale=0.9]{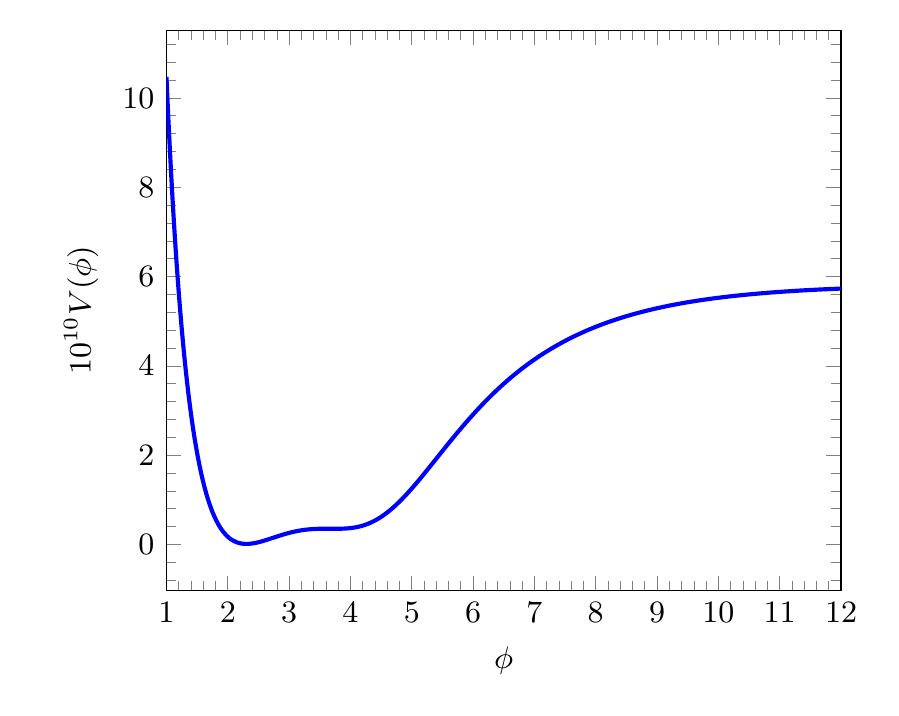}
      \end{minipage}
          \begin{minipage}{0.49\textwidth}
      \includegraphics[scale=0.9]{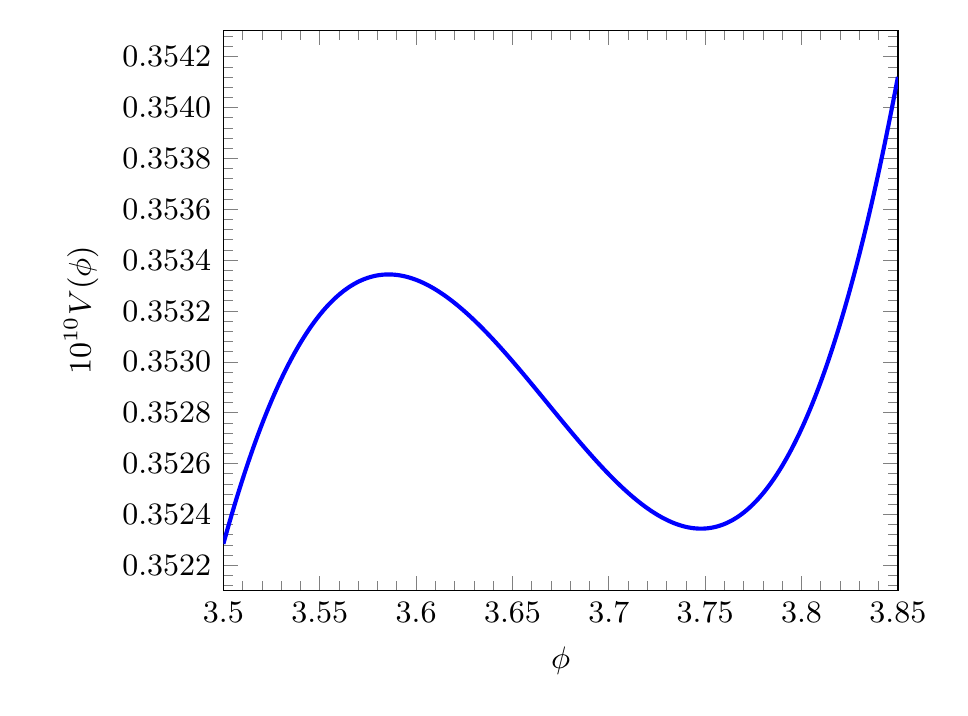}
      \end{minipage}
     \caption{\small Inflaton potential from Ref. \cite{sone3} with parameter set defined in Tab \ref{tab1}. On the right hand side, the detail of the local minimum and maximum around the inflection point.}      \label{figpot}
     \end{figure}

 We have checked that they provide the correct CMB normalisation at large scales, as well as the correct spectral index and PBH abundance to match the dark matter abundance.
The inflaton potential has  an inflection point violating the slow-roll conditions, see Fig. \ref{figpot}, where the field is forced to enter a non-attractor phase  which lasts a few e-folds and a boost in the  curvature power spectrum is generated.%
\noindent
The solution to the Friedmann equations shows that the background evolution can be divided as follows. In a first stage the field experiences a slow-roll evolution compatible with the constraints on CMB scales. When the field approaches the local minimum, the fields enters a non-attractor phase where $\eta =3$. After the following local maximum, the field exits the non-attractor phase leading to the end of the inflationary era.

\begin{figure}[ht!]
 \centering
      \begin{minipage}{0.49\textwidth}
      \includegraphics[scale=0.9]{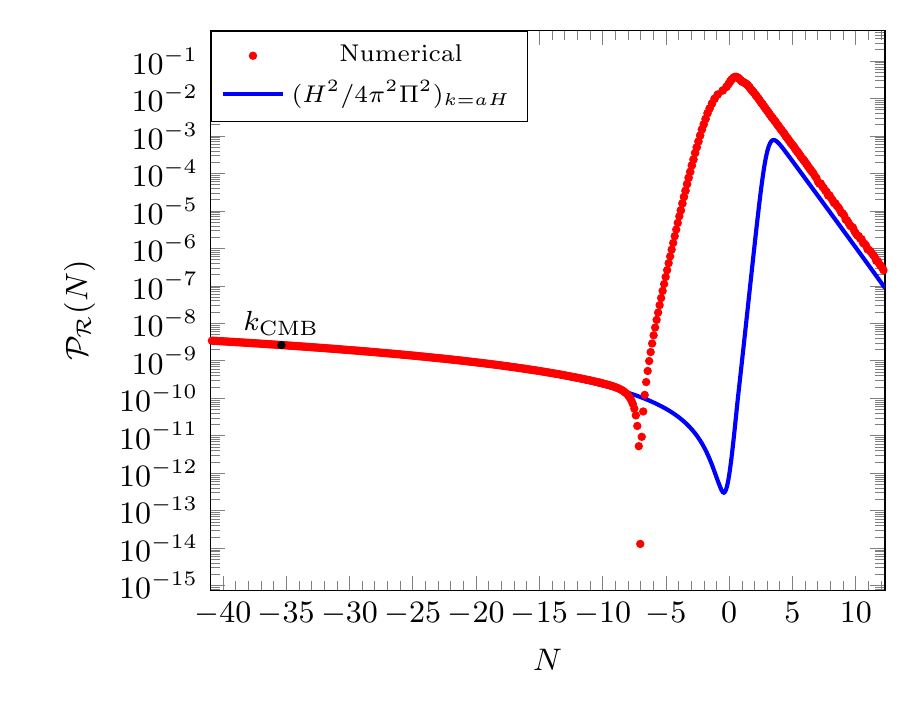}
      \end{minipage}
       \begin{minipage}{0.49\textwidth}
      \includegraphics[scale=0.9]{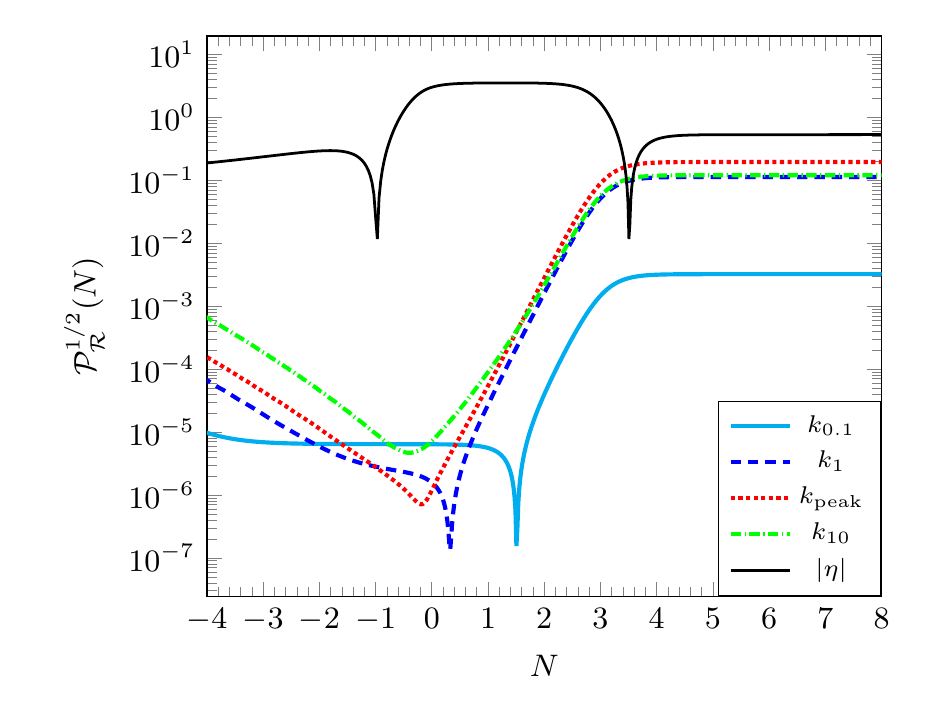}
      \end{minipage}
     \caption{\small The left figure shows the power spectrum for the model in  Ref. \cite{sone3} as a function of $N$ corresponding to $k=aH$ and where we have arbitrarily set $N=0$ at the time at which $\eta$ reaches 3. Together with the numerical solution of the power spectrum in the region of interest, the slow-roll formula is   plotted with the power spectrum computed at Hubble crossing. In the right figure the behaviour of $\eta$ and of the different modes are shown.  One can also observe that the transition of the parameter $\eta$ from its slow-roll value to 3 lasts $\sim 1$ e-fold. Notice that the modes
     stop growing when the non-attractor phase ends, that is when the $z^{-1}\dd z/\dd\tau$ becomes positive again. }      \label{figmodes}
\end{figure}

We numerically solve the equation for the comoving curvature perturbation $\mathcal{R} $ (cfr. Eq. \eqref{ff}) starting from the usual Bunch-Davies vacuum in the asymptotic region $(-k\tau) \gg 1$.
Fig. \ref{figmodes} shows the power spectrum and the behaviour of the different modes.
\noindent
We have indicated by $k_{\rm pk}$ the mode leading to the largest amplitude and  by $k_1$  the mode that  leaves the Hubble radius at the transition point (approximately at $N\simeq -1$ in Fig \ref{figmodes}). We have called    $k_{0.1}$ and $k_{10}$  the wavenumbers respectively 0.1 and 10 times larger than $k_1$. We have also plotted the 
power spectrum computed at Hubble crossing to show its inadequacy in reproducing the exact result which must be calculated at the end of the non-attractor phase. Notice also that the curvature perturbation ${\cal R}_{\rm pk}$ grows until the end of the non-attractor phase, meaning that it
takes advantage of the exponential decrease of the inflaton velocity until the end of the non-attractor phase, then it remains constant  until the end of inflation.
As we mentioned, this is because the transition back to the slow-roll phase is sudden and despite the fact that  the parameter $\eta$ does not go back immediately to very small values \cite{sasaki}. As for the absolute amplitude of the perturbation at the peak, we cannot really make use of  the formula (\ref{po}) since when the mode leaves the Hubble radius both $\epsilon$ and $\eta$ change 
considerably. Numerically, we have estimated ${\cal P}^{1/2}_{{\cal R}_{\rm pk}}\simeq 7(H/2\pi \Pi_\star)$. 

\begin{figure}[ht!]
 \centering
      \begin{minipage}{0.49\textwidth}
      \includegraphics[scale=0.9]{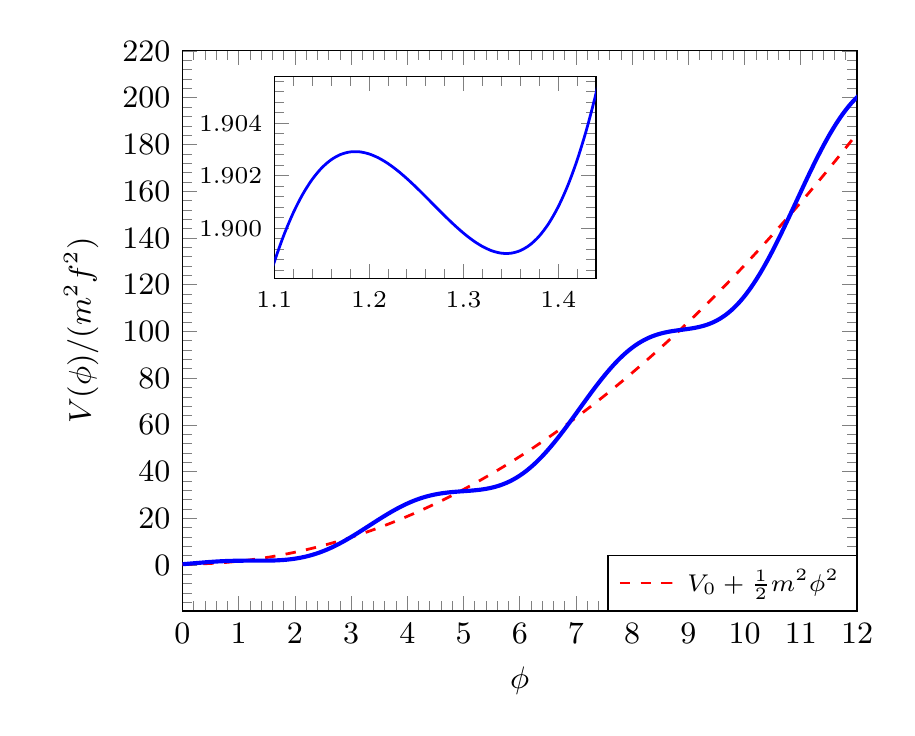}
      \end{minipage}
       \begin{minipage}{0.49\textwidth}
      \includegraphics[scale=0.9]{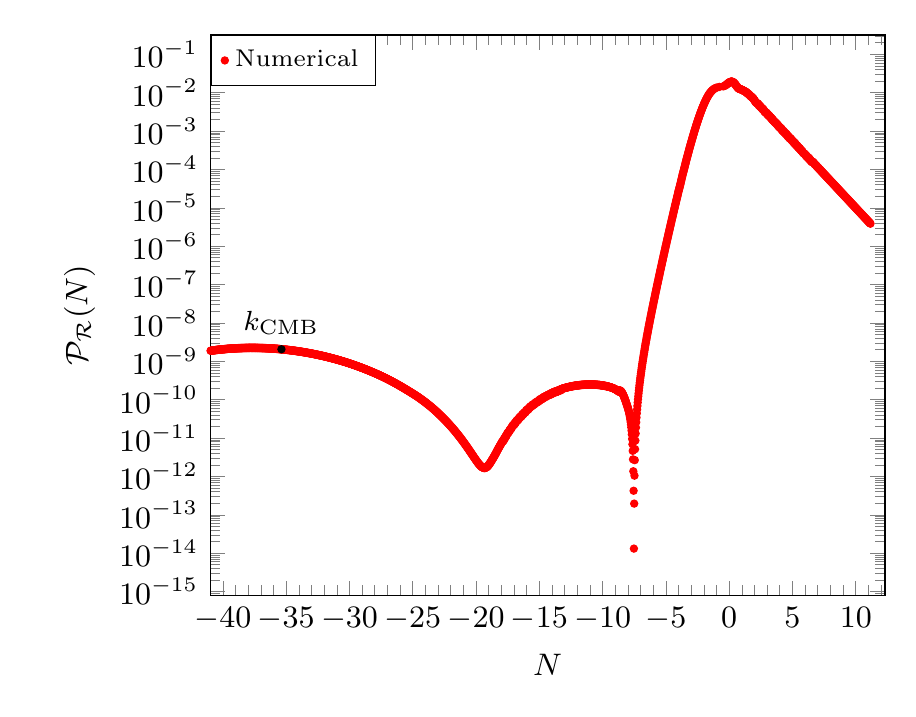}
      \end{minipage}
     \caption{\small On the left, the potential \eqref{tsp} of  Ref. \cite{sone4} compared to the quadratic potential. In the box, the detail of the potential near the inflection point. On the right, the corresponding power spectrum as a function of $N$ corresponding to $k=aH$ and where we have arbitrarily set $N=0$ at the time at which $\eta$ reaches 3.}      \label{figtas}
\end{figure}

In the following we will also perform our  analysis of the model in Ref. \cite{sone4}. The inflaton potential
\be\label{tsp}
V(\phi)=V_0 + \frac{1}{2} m^2 \phi^2 + \Lambda_1^4 \frac{\phi}{f} \cos\left(\frac{\phi}{f}\right)+\Lambda_2^4  \sin \left(\frac{\phi}{f}\right)
\ee
 is characterised by a series of oscillations around the quadratic potential, the last of which is capable of generating an inflection point, tuned such that the power-spectrum is enhanced as previously described for model \cite{sone3}, see Fig. \ref{figtas}.
\noindent
We have chosen to use the parameter set 1 in Ref. \cite{sone4} for our analysis for sake of comparison.  Also for this case we have numerically estimated  ${\cal P}^{1/2}_{{\cal R}_{\rm pk}}\simeq 7(H/2\pi \Pi_\star)$. The models in Refs. \cite{sone3,sone4} are similar to those of other recent literature \cite{sone1,sone0,sone2,ssm1} and we expect our analysis for the quantum diffusion to apply to those cases too.

\renewcommand{\theequation}{3.\arabic{equation}}
\setcounter{equation}{0}

\section*{III. The non-attractor phase and  quantum diffusion}
\noindent
Let us now come back to the role  of quantum diffusion. If too large,   quantum diffusion causes  a loss of information as the curvature perturbation may not  be reconstructed  any longer at late times in terms of classical trajectories \cite{sone2}. Different scales mix and the corresponding amplitude will be left  undetermined for an observer at late times. Since quantum diffusion becomes more and more relevant as the field slows down and consequently the power spectrum grows, this clearly creates an issue and one expects a upper bound on the curvature perturbation in order for the quantum diffusion to be irrelevant.
 
 Since the power spectrum is fixed by the inverse of the velocity of the inflaton field at the end of the non-attractor phase, we expect that,  if the spread of the distribution of velocities
 caused by the stochastic motion is too large, then along most of the trajectories the perturbation will be either too large or to  too small to generate PBHs in the allowed range of masses. One has therefore to find the amount of dispersion undergone by the velocity of the inflaton field.
 
 Let us also notice that the power spectrum is growing during the non-attractor phase after the corresponding wavelength
 leaves the Hubble radius and therefore the issue of the quantum diffusion becomes more relevant at the end of the non-attractor phase. 
 We will therefore discuss
 the criterion at the end of such a phase, where one expects the strongest constraints.

The stochastic equation (\ref{stoch}) can be written as an Ornstein-Uhlenbeck process
 \begin{eqnarray}
 \label{stochdue}
\frac{\dd \phi}{\dd N}&=&\Pi,\nonumber\\
 \frac{\dd \Pi}{\dd N}+3\Pi +\frac{V_{,\phi}}{H^2}&=&\xi,\nonumber\\
 \langle{\xi}(N){\xi}(N')\rangle&=&D\delta(N-N'),\nonumber\\
 D&=&\frac{9H^2}{4\pi^2},
 \end{eqnarray}
where $D$ is the diffusion coefficient.
We may write the  Kramers-Moyal (KM)  equation for the
corresponding probability 
$P(\phi,\Pi,N)$ as \cite{Risken}
\be
\label{q}
\frac{\partial P }{\partial N}=-\frac{\partial }{\partial \phi}\left(\Pi P\right)+
\frac{\partial}{\partial \Pi}\left[{\cal V}_{,\Pi}P+\frac{V_{,\phi}}{H^2}P\right]+\frac{D}{2}
\frac{\partial^2}{\partial \Pi^2}P,
\ee
where 
\be
\label{pot}
{\cal V}(\Pi)=\frac{3}{2}  \Pi^2.
\ee
The initial condition for the probability can be taken to be
\be
P(\phi,\Pi,0)=\delta_D(\phi-\phi_0)\delta_D(\Pi-\Pi_0),
\ee
 as we assume that during the
preceding slow-roll phase the motion is purely along classical trajectories. 

 \subsection*{Generic potential}
\noindent
 We re-write the KM equation as 
\begin{eqnarray}
\frac{\partial P}{\partial N}&=&L_{\rm KM}\, P=\left(L_{\rm rev}+L_{\rm ir}\right)\,P,\nonumber\\
L_{\rm rev}&=&-\Pi\frac{\partial}{\partial\phi}+\frac{V_{,\phi}}{H^2}\frac{\partial}{\partial\Pi},\nonumber\\
L_{\rm ir}&=&\frac{\partial}{\partial\Pi}\left(3\Pi+\frac{D}{2}\frac{\partial}{\partial\Pi}\right).\nonumber\\
\end{eqnarray}
The stationary solution for the operator $L_{\rm ir}$ is proportional to ${\rm exp}(-3\Pi^2/D)$ and we can generate a Hermitian operator
\be
\overline{L}_{\rm ir}={\rm exp}\left(\frac{3\Pi^2}{2D}\right) L_{\rm ir}\,{\rm exp}\left(-\frac{3\Pi^2}{2D}\right)=\overline{L}_{\rm ir}^\dagger=-3a^\dagger a,
\ee
where 
\be
a=\sqrt{\frac{D}{6}}\frac{\partial}{\partial\Pi}+\frac{\Pi}{2}\sqrt{\frac{6}{D}},\,\,\,\,a^\dagger=-\sqrt{\frac{D}{6}}\frac{\partial}{\partial\Pi}+\frac{\Pi}{2}\sqrt{\frac{6}{D}}
\ee
are the annihilation and creator operators with $[a,a\dagger]=1$. To take advantage of this procedure, we also redefine the operator
\be
\overline{L}_{\rm rev}={\rm exp}\left(\frac{3\Pi^2}{2D}+\rho\frac{V}{H^2}\frac{6}{D}\right) L_{\rm rev}\,{\rm exp}\left(-\frac{3\Pi^2}{2D}-\rho\frac{V}{H^2}\frac{6}{D}\right) =-a A-a^\dagger \hat A,
\ee
where $\rho$ is an arbitrary constant 
\begin{eqnarray}
A=\sqrt{\frac{D}{6}}\frac{\partial}{\partial\phi}-\rho\frac{V}{H^2}\sqrt{\frac{6}{D}},\,\,\,\, \,\,\,\,\hat A=\sqrt{\frac{D}{6}}\frac{\partial}{\partial\phi}+(1-\rho)\frac{V}{H^2}\sqrt{\frac{6}{D}},
\end{eqnarray}
with $[A,\hat A]=V_{\phi\phi}/H^2$.
Now, the orthonormalised eigenfunctions of the operator $\overline{L}_{\rm ir}$, that is
\be
\overline{L}_{\rm ir}\phi_n(\Pi)=-3n\phi_n(\Pi),\,\,\,\, a^\dagger a\phi_n(\Pi)=n\phi_n(\Pi),
\ee
are
\be
\phi_n(\Pi)=(a^\dagger)^n \phi_0(\Pi)/\sqrt{n!},\,\,\,\,\phi_0(\Pi)=\frac{{\rm exp}\left(-3\Pi^2/2D\right)}{\sqrt{D/6\sqrt{2\pi}}}.
\ee
Since the operator $L_{\rm KM}$ is of the form
\be
L_{\rm KM}={\rm exp}\left(-\frac{3\Pi^2}{2D}-\rho\frac{V}{H^2}\frac{6}{D}\right)\left(\overline{L}_{\rm ir}+\overline{L}_{\rm rev}\right)\,{\rm exp}\left(\frac{3\Pi^2}{2D}+\rho\frac{V}{H^2}\frac{6}{D}\right)\phi^{-1}_0(\Pi),
\ee
or
\be
L_{\rm KM}=-\phi_0(\Pi){\rm exp}\left(-\rho\frac{V}{H^2}\frac{6}{D}\right)\left(3a^\dagger a+aA+a^\dagger \hat A\right){\rm exp}\left(\rho\frac{V}{H^2}\frac{6}{D}\right)\phi^{-1}_0(\Pi),
\ee
we can expand the probability as \cite{Risken}
\be
P=\phi_0(\Pi){\rm exp}\left(-\rho\frac{V}{H^2}\frac{6}{D}\right)\sum_{n\geq 0}c_n(\phi,N)\phi_n(\Pi),
\ee
so that the distribution in the inflaton field is only given by the first term of the expansion
\be
\int\dd\Pi \,P={\rm exp}\left(-\rho\frac{V}{H^2}\frac{6}{D}\right)c_0(\phi,t),
\ee
where nevertheless the coefficients $c_n$ satisfy the so-called Brinkman's hierarchy
\be
\frac{\partial c_n}{\partial N}=-\sqrt{n} \hat A  c_{n-1}-3n c_n-\sqrt{n+1}Ac_{n+1}
\ee
and it is equivalent to the KM equation. This equation contains an  infinite number of terms. For pedagogical purposes,  
let us truncate  though the system by setting $c_n=0$ for $n\geq 3$, so that the Brinkman's hierarchy reduces to
\begin{eqnarray}
\frac{\partial c_0}{\partial N}+Ac_{1}&=&0,\nonumber\\
\frac{\partial c_1}{\partial N}+\hat A c_0+3 c_1&=&0.
\end{eqnarray}
For a large friction term one can neglect the term $\partial c_1/\partial N$, and we could eliminate $c_1$ in favour of $c_0$. Setting $\rho=0$, we find
\be
\frac{\partial c_0}{\partial N}=-A c_1=\frac{1}{3}A\hat A c_0=
\frac{1}{3H^2}\frac{\partial}{\partial\phi}\left(V_{,\phi} c_0\right)+\frac{1}{2}\frac{D}{9}\frac{\partial^2 c_0}{\partial\phi^2},
\ee
which is the standard Fokker-Planck equation. Had not we dropped the term $\partial c_1/\partial N$, we could have eliminated $c_1$ and get the equation for $c_0$
\be
\frac{\partial^2 c_0}{\partial N^2}+3\frac{\partial c_0}{\partial N}=
\frac{1}{H^2}\frac{\partial}{\partial\phi}\left(V_{,\phi} c_0\right)+\frac{D}{6}\frac{\partial^2 c_0}{\partial\phi^2},
\ee
which is Brinkman's equation. 
Retaining the coefficients $c_n$ with $n\geq 3$ will introduce spatial derivatives higher than two. We find  here what we mentioned in the introduction, that the KM contains  an infinite tower of spatial
derivatives of the  effective probability of the inflaton field and, due to Pawula's theorem,  it  is not consistent to drop derivatives higher than two. In this sense, the Fokker-Planck equation is not the correct starting point. 

In the case of a linear potential 
 the operators $A$ and $\hat A$ commute, while for a quadratic potential their commutator is a constant and the analysis is made it easier. We will consider these cases next.
 \subsection*{Linear potential}
\noindent
We consider first   the linear potential (\ref{linear}) as a prototype. In such a case, the KM equation has an exact solution  \cite{Risken}
 \begin{eqnarray}
 \label{km}
 P(\phi,\Pi,N)&=&\frac{1}{2\pi \left({\rm Det}{\bf M}\right)^{1/2}}{\rm exp}\left\{-1/2[{\bf M}^{-1}]_{\phi\phi}(\Delta \phi)^2-[{\bf M}^{-1}]_{\phi\Pi}\Delta \phi
\Delta \Pi- 1/2[{\bf M}^{-1}]_{\Pi\Pi}(\Delta \Pi)^2\right\},\nonumber\\
&&
 \end{eqnarray}
where  
\begin{eqnarray}
\Delta\phi&=&\phi-\phi(N),\nonumber\\
 \Delta\Pi&=&\Pi-\Pi(N),\nonumber\\
\langle\phi(N)\rangle&=&\phi(N)=\phi_0+\frac{1}{3}(\Pi_0-\Pi)-\sqrt{2\epsilon_V}N,\nonumber\\
\langle\Pi(N)\rangle&=&\Pi(N)=\sqrt{2\epsilon_V}\left(e^{-3N}-1\right)+\Pi_0 e^{-3N},
\end{eqnarray}
and 
\begin{eqnarray}
{\bf M}_{\phi\phi}&=&
\frac{D}{54}\left(6N-3+4 e^{-3N}-e^{-6N}\right),\nonumber\\
{\bf M}_{\phi\Pi}&=&\frac{D}{18}\left(1-e^{-3N}\right)^2,\nonumber\\
{\bf M}_{\Pi\Pi}&=&\frac{D}{6}\left(1-e^{-6N}\right).
\end{eqnarray}
At  times  $N\gsim 1$, the probability  becomes  
\be
P(\phi,\Pi,N)=\frac{1}{\Pi}\left(\frac{27}{2 D^2 N}\right)^{1/2}{\rm exp}\left[-\frac{9 }{2 D N} (\Delta\phi)^2\right]{\rm exp}\left[\frac{3}{ D N}
\Delta\phi\Delta\Pi\right]
{\rm exp}\left[-\frac{3 }{D}(\Delta\Pi)^2\right].
\ee
Integrating over $\Pi$ we  obtain
\be
P_{\phi}(\phi,N)=\frac{3}{\sqrt{2 \pi D N}}\,{\rm exp}
\left[-\frac{9 }{2 D N}(\phi-\phi(N))^2\right]
\ee
and
\be
\label{o}
\langle(\Delta\phi)^2\rangle =\int\,\dd\phi\,(\phi-\phi(N))^2\,P_{\phi}(\phi,N)=\frac{D}{9}N.
\ee
Conversely, integrating over the scalar field $\phi$, one obtains
\be
\label{pp0}
P_{\Pi}(\Pi,N)=\sqrt{\frac{3}{\pi D}}\,{\rm exp}
\left[-\frac{3 }{ D }(\Pi-\Pi(N))^2\right]
\ee
and
\be
\label{oo}
\langle(\Delta\Pi)^2\rangle =\int\,\dd\Pi\,(\Pi-\Pi(N))^2\,P_{\Pi}(\Pi,N)=\frac{D}{6 }.
\ee
We conclude that  when the average velocity of the inflaton field decays exponentially, its variance reaches quickly
an asymptotic and stationary value  given by Eq. (\ref{oo}).
There is an alternative way to  obtain the same result.  From the KM equation, we may
derive the following set of equations 
\begin{eqnarray}
\frac{\partial}{\partial N}\langle (\Delta\phi)^2\rangle&=&2\langle\Delta\phi \Delta\Pi\rangle\, ,\nonumber\\
\frac{\partial}{\partial N}\langle \Delta\phi \Delta\Pi\rangle&=&\langle (\Delta\Pi)^2\rangle
-3\langle\Delta\phi \Delta\Pi\rangle\, ,\nonumber\\
\frac{\partial}{\partial N}\langle (\Delta\Pi)^2\rangle&=&-6\langle (\Delta\Pi)^2\rangle
+D.
\end{eqnarray}
At times larger than a few Hubble times, the correlators involving $\Delta\Pi$ decay promptly
to their equilibrium values $\langle\Delta\phi \Delta\Pi\rangle=1/3\langle (\Delta\Pi)^2\rangle=D/18$ resulting in 
\be
\frac{\partial}{\partial N}\langle (\Delta\phi)^2\rangle=\frac{D}{9},
\ee
reproducing  (\ref{o}) and (\ref{oo}).

 \subsection*{Linear plus quadratic potential}
\noindent
Our considerations can be extended  beyond the linear order in the potential. Let us  expand the potential  including the quadratic order 
\be
V(\phi)=V_0\left[1+\sqrt{2\epsilon_V}(\phi-\phi_0)+\frac{1}{2}\eta_V(\phi-\phi_0)^2\right]+\cdots, 
\ee
where $\eta_V=V_{,\phi\phi}/3H^2$ parametrises the  second derivative of the potential. 
The equation of motion leads to a classical value
\be
\langle \Pi(N)\rangle=\Pi(N)\simeq \Pi_0e^{-3N}\left(1+\frac{1}{3}\eta_V+\frac{\sqrt{2\epsilon_V}}{\Pi_0}\right)
-\Pi_0e^{-\eta_VN}\left(\frac{1}{3}\eta_V+\frac{\sqrt{2\epsilon_V}}{\Pi_0}\right).
\ee
In particular, if one has a potential where $\phi_0$ corresponds to a minimum and only the quadratic piece is there in the Taylor expansion
one finds
\be
\langle \Pi(N)\rangle=\Pi(N)\simeq \Pi_0e^{-3N}\left(1+\frac{1}{3}\eta_V\right)
-\frac{\Pi_0}{3}\eta_V e^{-\eta_VN}.
\ee
In order to simplify the problem,  we notice that in the stochastic equation of motion of the inflaton field 
 \begin{eqnarray}
 \frac{\dd^2 \phi}{\dd N^2}+3\frac{\dd \phi}{\dd N} +3\sqrt{2\epsilon_V}+3\eta_V(\phi-\phi_0)&=&\xi,
 \end{eqnarray}
one can shift the field $\Pi$ by an amount $-3\sqrt{2\epsilon_V}$ and $\phi$ by an amount $-3\sqrt{2\epsilon_V}N$  in order to get rid of the constant force. The problem reduces for this shifted field
to the following set of equations (we do not redefine the fields to avoid cluttering notation)
 \begin{eqnarray}
\frac{\dd \phi}{\dd N}&=&\Pi,\nonumber\\
 \frac{\dd \Pi}{\dd N}+3\Pi +3\eta_V\phi&=&\xi.
 \end{eqnarray}
 The solution of these equations is again given in Eq. (\ref{km}). This time however
 \begin{eqnarray}
\phi(N)&=&[{\rm exp}(-{\bf A}N)]_{\phi\phi}\phi_0+[{\rm exp}(-{\bf A}N)]_{\phi\Pi}\Pi_0,\nonumber\\
\Pi(N)&=&[{\rm exp}(-{\bf A}N)]_{\Pi\phi}\phi_0+[{\rm exp}(-{\bf A}N)]_{\Pi\Pi}\Pi_0,\nonumber\\
{\bf A}&=&\left(\begin{array}{cc}
0  & -1\\
3\eta_V & 3
\end{array}\right).
\end{eqnarray}
 Also, by defining
 \be
 \lambda_{1,2}=\frac{1}{2}\left(3\pm\sqrt{9-12\eta}\right),\,\,\,\,\lambda_1\simeq 3,\,\,\,\,\lambda_2\simeq \eta_V,
 \ee
 one obtains \cite{Risken}
 \begin{eqnarray}
{\bf M}_{\phi\phi}&=&
\frac{D}{2(\lambda_1-\lambda_2)^2}\left[\frac{\lambda_1+\lambda_2}{\lambda_1\lambda_2}+\frac{4}{\lambda_1+\lambda_2}\left(e^{-(\lambda_1+\lambda_2)N}-1\right)
-\frac{1}{\lambda_1}e^{-2\lambda_1N}-\frac{1}{\lambda_2}e^{-2\lambda_2N}\right],\nonumber\\
{\bf M}_{\phi\Pi}&=&\frac{D}{2(\lambda_1-\lambda_2)^2}\left(e^{-\lambda_1N}-e^{-\lambda_2N}\right)^2,\nonumber\\
{\bf M}_{\Pi\Pi}&=&\frac{D}{2(\lambda_1-\lambda_2)^2}\left[\lambda_1+\lambda_2+\frac{4\lambda_1\lambda_2}{\lambda_1+\lambda_2}\left(e^{-(\lambda_1+\lambda_2)N}-1\right)
-\lambda_1 e^{-2\lambda_1 N}-\lambda_2 e^{-2\lambda_2 N}\right],
\end{eqnarray}
or
\begin{eqnarray}
{\bf M}_{\phi\phi}&=&
\frac{D}{18}\left[\frac{1}{3}+\frac{1}{\eta_V}+\frac{4}{3}\left(e^{-3N}-1\right)
-\frac{1}{3}e^{-6N}-\frac{1}{\eta_V}e^{-2\eta_VN}\right],\nonumber\\
{\bf M}_{\phi\Pi}&=&\frac{D}{18}\left(e^{-3N}-e^{-\eta_V N}\right)^2,\nonumber\\
{\bf M}_{\Pi\Pi}&=&\frac{D}{18}\left[3+4\eta_V\left(e^{-3N}-1\right)
-3 e^{-6 N}-\eta_V e^{-2\eta_V N}\right].
\end{eqnarray}
At large times and for small $\eta_V$ they reduce
to
\begin{eqnarray}
{\bf M}_{\phi\phi}&=&
\frac{D}{18}\left[\frac{1}{\eta_V}\left(1-e^{-2\eta_VN}\right)-1
\right],\nonumber\\
{\bf M}_{\phi\Pi}&=&\frac{D}{18}e^{-2\eta_V N},\nonumber\\
{\bf M}_{\Pi\Pi}&=&\frac{D}{18}\left(3
-\eta_V e^{-2\eta_V N}\right).
\end{eqnarray}
%
%
 For  $\eta_V>0$, i.e. for a harmonically bound state,  in the large time limit one obtains a stationary solution. 
 However,  for  $\eta_V<0$, i.e. for an inverted parabolic potential,  the force felt by the inflaton is repulsive.  In both cases, the width of the distribution of the inflaton velocities obtained integrating out over all possible values of the inflaton field reads
\be\label{refqua}
\langle(\Delta\Pi)^2\rangle\simeq 
\frac{D}{18}\left(3-\eta_Ve^{-2\eta_V N}\right).
\ee 
In the  model of Ref. \cite{sone3,sone4}  the plateau is in fact a region around an inflection point between a minimum and a maximum so that  $\eta_V$ changes sign 
from positive to negative (if the minimum is encountered first). Being the dynamics more complex than what described above, we should   expect  deviations  of  order unity from our estimate.
\renewcommand{\theequation}{4.\arabic{equation}}
\setcounter{equation}{0}

 \section*{IV. Numerical analysis of quantum diffusion}
\noindent
In this section we present the numerical studies we performed in order to check the validity of  our analytical findings. We have numerically solved the system \eqref{stochdue} with the available Mathematica routines for the solution of stochastic differential equations.  We focus only on the inflaton velocity since the perturbations
are sensitive to it. The spread in the inflaton field, which acquires typically Planckian values (at least in the vast majority of the literature)  is irrelevant. At any rate, we have numerically checked that our numerical results coincide with this statement.  We were able to test the robustness of our numerical implementation for the case of the linear potential, for which we have the analytical solution, Eq. \eqref{km}.

\subsection*{Linear potential and Starobinsky's model}
\noindent
We start by checking the solution of the KM equation in the case of a linear potential.  Since we are interested in the dispersion of $\Pi_\star$ around its classical value, we recover numerically its variance among many realizations of the stochastic evolution.

\begin{figure}[ht!]
 \centering
 \hspace{-1cm}
    \begin{minipage}{0.49\textwidth}
      \includegraphics[scale=0.9]{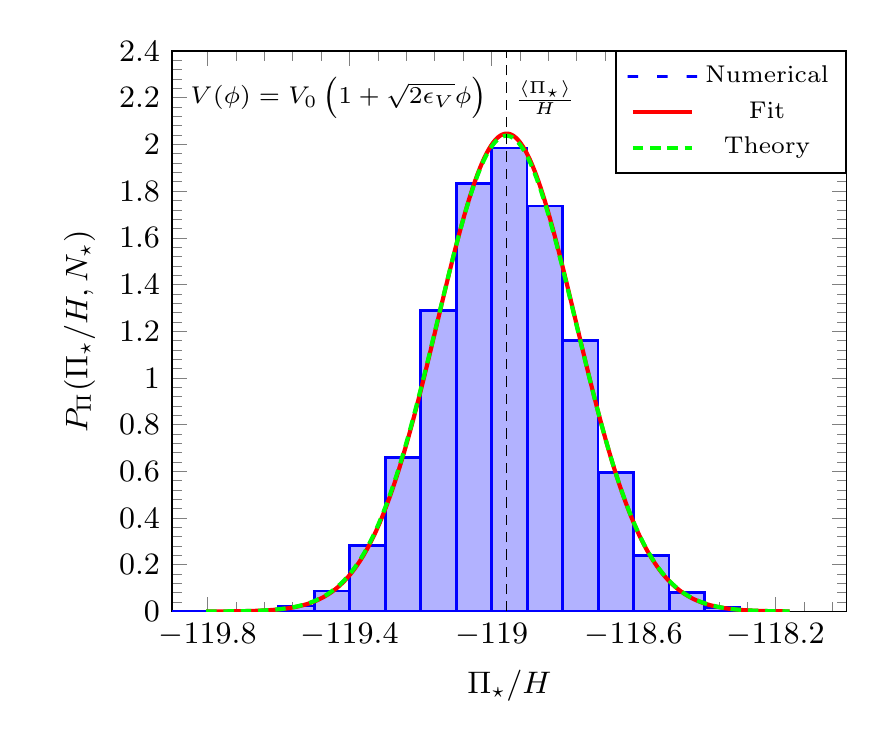}
      \end{minipage}
          \begin{minipage}{0.49\textwidth}
      \includegraphics[scale=0.9]{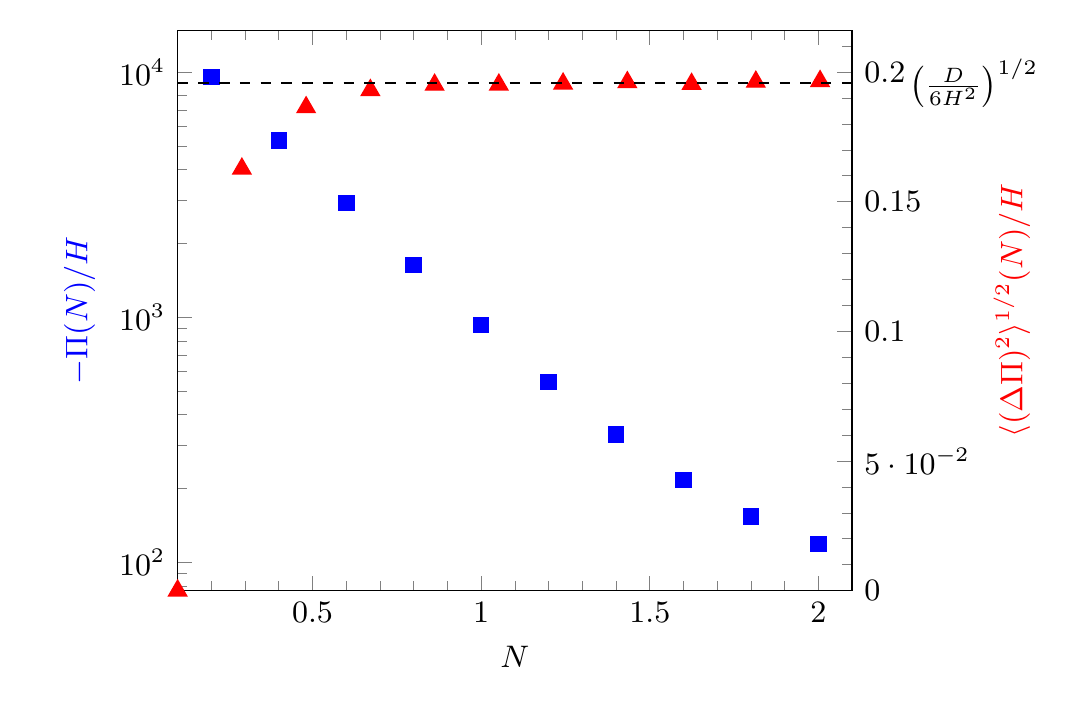}
      \end{minipage}
     \caption{\small  On the left, the numerical, the fit to numerical and the theoretical prediction for the probability $P_\Pi(\Pi_\star,N_\star)$ for the case of a 
     linear potential with $\epsilon_V=10^{-7}$. On the right, the classical evolution of $\Pi (N)$ together with the spread $\langle(\Delta\Pi)^2\rangle^{1/2}$ which stabilises at $(D/6)^{1/2}$ for $N\gtrsim 1$. It was checked numerically that changing the initial condition for $\Pi (0)$ by order of magnitudes does not give rise to significant modification of $\langle(\Delta\Pi)^2\rangle^{1/2}$ for $N<1$. Furthermore, the plateau's value is not sensitive to the initial conditions, confirming the analytical result in Eq. \eqref{oo}. }      \label{fig1}
\end{figure}

\begin{figure}[ht!]
 \centering
 \hspace{-1cm}
    \begin{minipage}{0.49\textwidth}
      \includegraphics[scale=0.9]{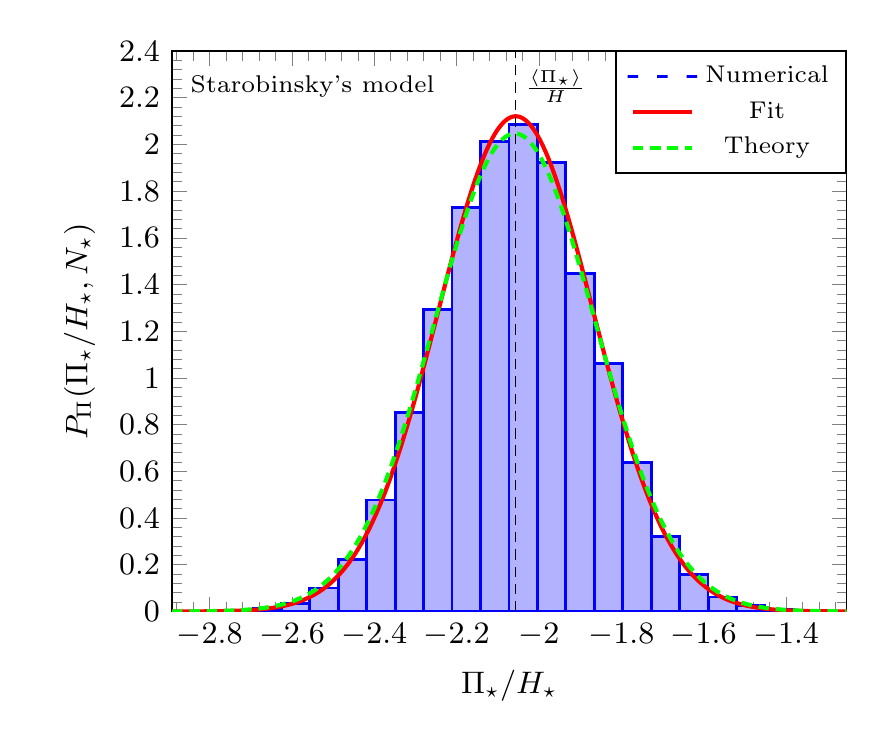}
      \end{minipage}
       \hspace{-0cm}
          \begin{minipage}{0.49\textwidth}
      \includegraphics[scale=0.9]{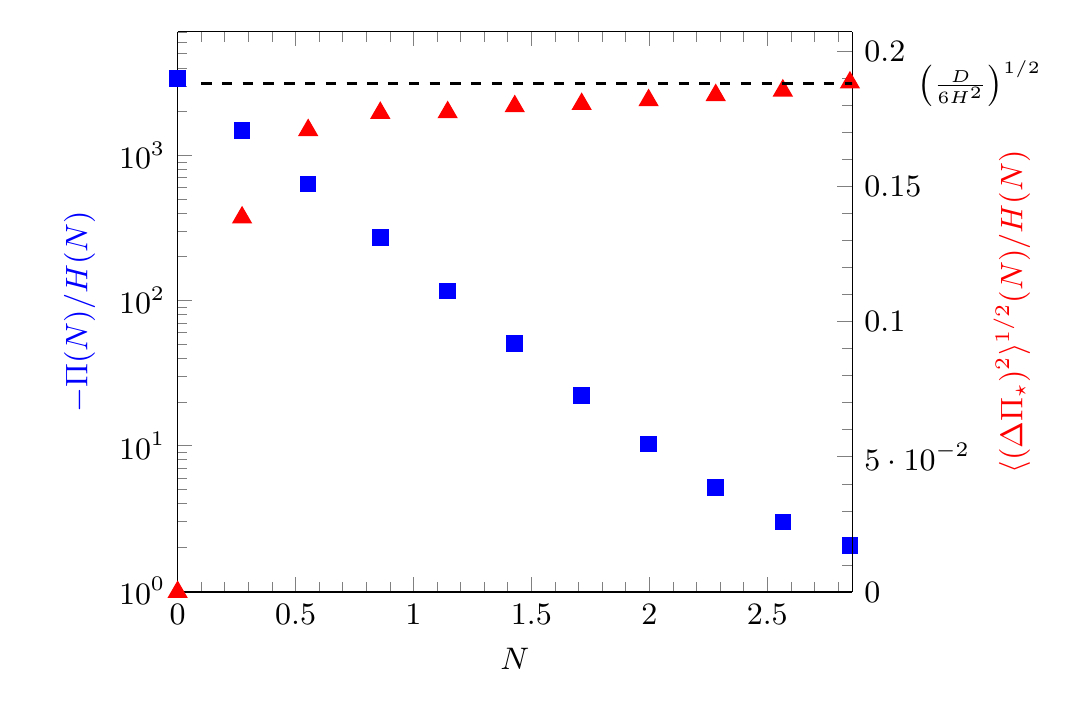}
      \end{minipage}
     \caption{\small  The numerical, the fit to numerical and the theoretical prediction for the probability $P_\Pi(\Pi_\star,N_\star)$ for the case of 
  Starobinsky's model with $\delta=0.01$ and $\epsilon_+/\epsilon_-=10^8$. On the right, the classical evolution of $\Pi (N)$ together with the spread $\langle(\Delta\Pi)^2\rangle^{1/2}$ which tends towards $(D/6)^{1/2}$. We have used the same parameters as in section II and performed $5\cdot 10^4$ realisations of the stochastic evolution.}      
  \label{figsn}
\end{figure}

In Fig. \ref{fig1}, one can see the comparison between the prediction \eqref{oo} and the numerical results. The numerical
results, obtained integrating over the inflaton field positions (whose spread is however tiny with respect to the average classical position),  fully reproduce the analytical results (to the extent that the red line of the fit and the green one representing the theory overlap perfectly).
\noindent
\noindent
%
%
We have also repeated our analysis for Starobinsky's model \cite{Starobinsky92} we have introduced in section II. The results are in Fig. \ref{figsn} which show that the  spread
of the velocity approaches  $(D/6)^{1/2}$. For smaller values of $\delta$, the agreement with the linear potential result would be extended to the whole non-attractor phase, but the choice of $\delta$ is limited by numerical precision.



\noindent

\subsection*{More physical cases}
\noindent
Having established that the numerical and analytical results agree for the simple case of the linear potential, we now
 turn our attention to more realistic cases discussed in the literature. As a representative example of the models in the literature, we consider the ones described in Ref. \cite{sone3} and \cite{sone4} already introduced in Sec. II. We solve numerically the stochastic equations (\ref{stochdue}) setting up initial conditions deep enough in the slow-roll phase. It was checked that, as expected, the stochastic noise can be neglected throughout the entire slow-roll phase. We focus our attention on the dynamics during the non-attractor phase. 

\begin{figure}[ht!]
 \centering
 \hspace{-1cm}
    \begin{minipage}{0.49\textwidth}
      \includegraphics[scale=0.9]{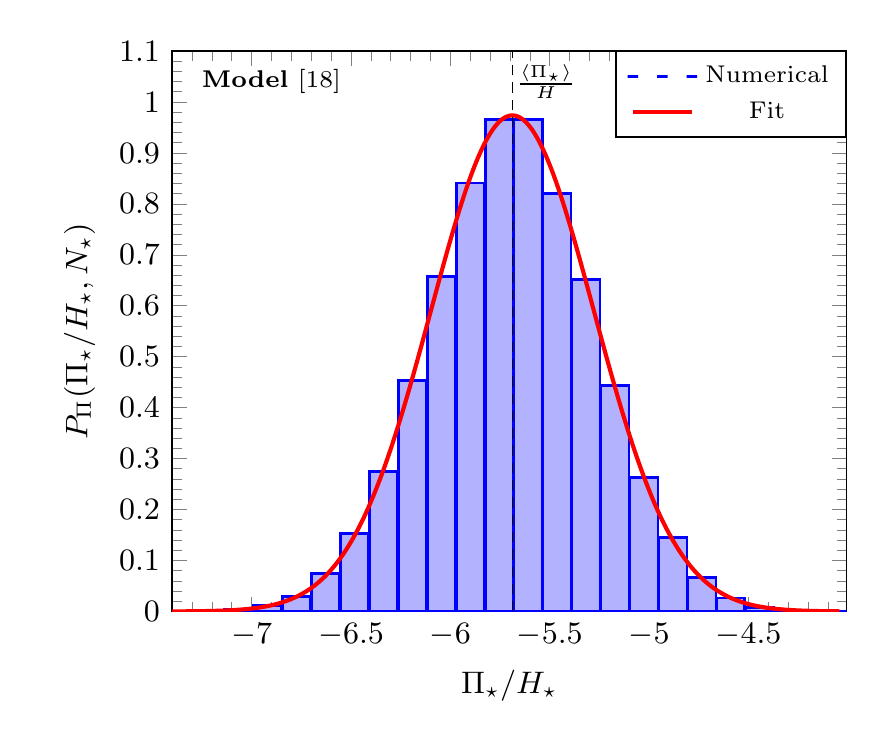}
      \end{minipage}
     \hspace{0.4cm}
          \begin{minipage}{0.49\textwidth}
      \includegraphics[scale=0.9]{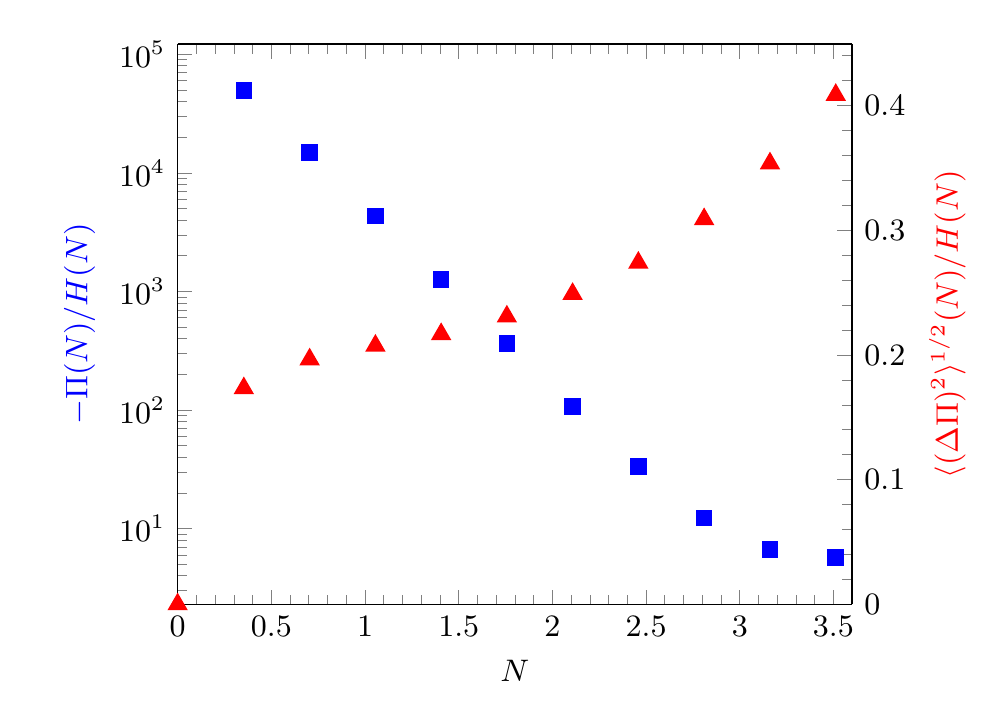}
      \end{minipage}
     \caption{\small  On the left the velocity probability obtained numerically together with its fit. On the right the evolution with time of the classical value of the velocity and its
     spread during the non-attractor phase. Calculations done for the model in Ref. \cite{sone3} with the parameter set in Tab. \ref{tab1}. The results are based on $5\cdot 10^4$ realisations of the stochastic evolution.}      \label{fig1.2}
\end{figure}

\begin{figure}[ht!]
 \centering
 \hspace{-1cm}
    \begin{minipage}{0.49\textwidth}
      \includegraphics[scale=0.9]{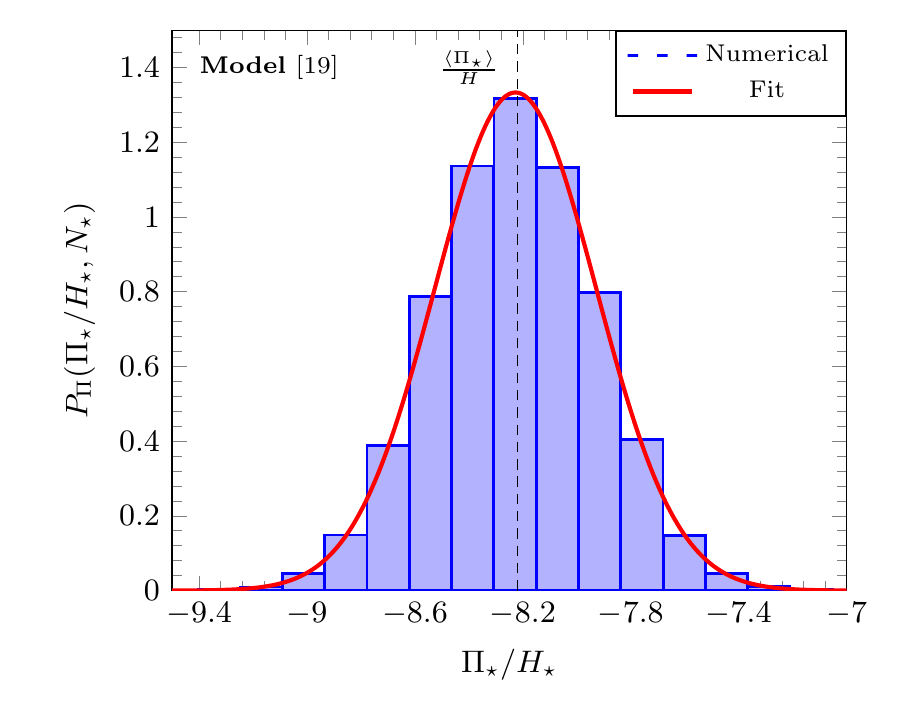}
      \end{minipage}
     \hspace{0.4cm}
          \begin{minipage}{0.49\textwidth}
      \includegraphics[scale=0.9]{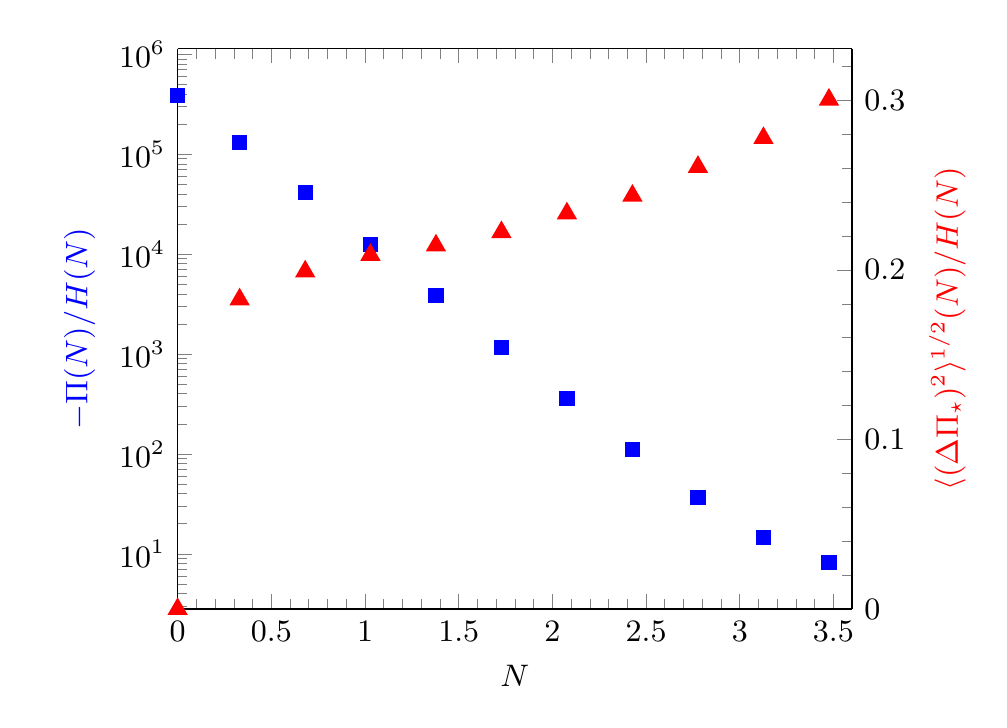}
      \end{minipage}
     \caption{\small  Same as Fig. \ref{fig1.2}, but for Ref. \cite{sone4}, case 1. }      \label{fig1.2.tasinato}
\end{figure}

In Fig. \ref{fig1.2} and Fig. \ref{fig1.2.tasinato} one can observe the evolution of $\Pi(N)$ and its dispersion around the mean value along the non-attractor phase, where the number of e-folds is set to zero at the transition. The procedure followed for the marginalisation over the $\phi(N)$ field is the same as the one previously presented for the case of a linear potential. 
\noindent
\noindent

We see that the more the classical value of the inflaton velocity decreases, the more its spread  grows with time. 
The distribution is well-fitted by a Gaussian with spread $\langle(\Delta\Pi)^2\rangle(N)$. In the former model, for example, at the end of the non-attractor phase,  we have 
\be
\frac{\langle(\Delta\Pi)^2\rangle^{1/2}}{H}\simeq 0.4,
\ee
which is larger than about a factor of two than the variance for  the linear potential $\sqrt{D/6}/H\simeq 0.2$. We notice instead that the behaviour of the spread is well reproduced by the expression (\ref{refqua}) even though with deviations near the end of the non-attractor phase.


\renewcommand{\theequation}{5.\arabic{equation}}
\setcounter{equation}{0}

  \section*{V. A  criterion for the quantum diffusion}
\begin{figure}[ht!]
    \begin{center}
      \includegraphics[scale=.9,angle=360]{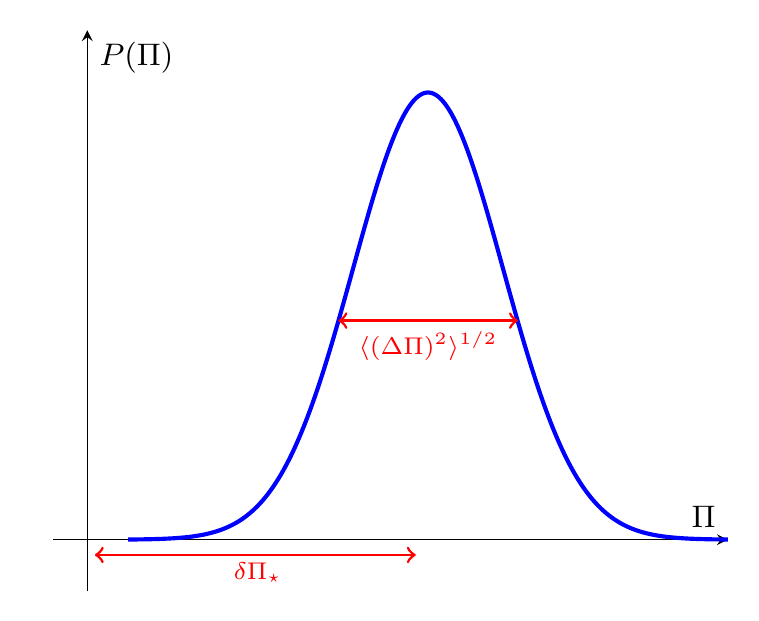}
    \end{center}
     \caption{\small A representative behaviour of the quantum diffusion issue. The spread of the inflaton velocity probability $\langle(\Delta\Pi)^2\rangle^{1/2}$ has to be smaller than
     the distance between the origin and the average value of the velocity.}      \label{one1}
\end{figure}

\noindent
As previously discussed, the crucial quantity is the spread of the velocity $\Delta\Pi$ of the inflaton field for the various trajectories. 
If the spread of the probability distribution $\langle(\Delta\Pi)^2\rangle^{1/2}$ is smaller than the  size $\delta\Pi_\star$ of the region  over which  the perturbation is of the order of  ${\cal P}^{1/2}_{{\cal R}_{\rm pk}}$, then an insignificant part of the 
wave packet goes out  the region where the curvature
perturbation is ${\cal P}^{1/2}_{{\cal R}_{\rm pk}}$ and  most of the trajectories will have the same  curvature
perturbation  $\sim  {\cal P}^{1/2}_{{\cal R}_{\rm pk}}$.  We impose therefore the criterion  that the spread of the probability distribution  is still within the region where ${\cal P}^{1/2}_{\cal R}\sim  {\cal P}^{1/2}_{{\cal R}_{\rm pk}}$, see Fig. \ref{one1},
\noindent
\be
\boxed{
\frac{\langle (\Delta\Pi)^2\rangle^{1/2}}{H} \ll \frac{\delta\Pi_\star}{H}.}
\ee
\subsection*{Linear potential}
\noindent
For the linear potential during the non-attractor phase,  the region where the curvature perturbation has a given value  ${\cal P}^{1/2}_{{\cal R}_{\rm pk}}$  has a width
(recall that   ${\cal P}_{{\cal R}_{\rm pk}}\sim 2.5{\cal P}_{{\cal R}_{\star}}$)
\be
\delta\Pi_\star\simeq 1.6\frac{H}{2\pi  {\cal P}^{1/2}_{{\cal R}_{\rm pk}}}.
\ee
and therefore one obtains the criterion
\be
\label{loo}
 {\cal P}^{1/2}_{{\cal R}_{\rm pk}}\ll 1.6\frac{H}{2\pi\langle (\Delta\Pi)^2\rangle^{1/2}}.
\ee
If satisfied, we can conclude that along most of the classical trajectories PBH's can be generated. If not true, this is equivalent to say that
the wave-function of the inflaton velocity penetrates into the regions where the velocities are much different from $\Pi_\star$, leading to   non-perturbative values of ${\cal R}$ and to a totally
random motion if $\Pi_\star\sim 0$. Of course, one can get a stronger constraint if one imposes that the penetration does not occur at $p$ variances, or
\be
 {\cal P}^{1/2}_{{\cal R}_{\rm pk}}\ll 1.6 \frac{H}{2p\pi\langle (\Delta\Pi)^2\rangle^{1/2}}.
\ee
Now, since 
\be
\sqrt{\frac{D}{6}}\ll \delta\Pi_\star,
\ee
we finally obtain
\be
\label{lo}
 {\cal P}^{1/2}_{{\cal R}_{\rm pk}}\ll 1.6\sqrt{\frac{2}{3p^2}}\simeq \frac{1.3}{p}.
\ee
\subsection*{More physical cases}
\noindent
For the more realistic models discussed in Refs. \cite{sone3,sone4}  our results are provided in Figs. \ref{fig1.2} and \ref{fig1.2.tasinato}. As we have noticed, 
the distribution is well-fitted by a Gaussian with spread $\langle(\Delta\Pi)^2\rangle(N)$. As already mentioned, one needs to take the value of the curvature perturbation at the peak at the end of the non-attractor phase since the corresponding mode does not change in time afterwards. Therefore, taking into account that for both cases ${\cal P}^{1/2}_{{\cal R}_{\rm pk}}\simeq 7(H/2\pi \Pi_\star)$, we obtain
\be
\label{w}
\frac{\delta\Pi_\star}{H}\simeq 1.4,
\ee
while 
\be
\frac{\langle(\Delta\Pi)^2\rangle^{1/2}}{H}\simeq 0.4,
\ee
which is  comfortably smaller than (\ref{w}). The criterion is well satisfied thanks to the boost the power spectrum gets at the peak with respect to the power spectrum calculated
for the wavelength leaving the Hubble radius deep in the non-attractor phase. However, as we will see next, this does not seem enough for the quantum diffusion not to have an impact on the PBH abundance.

\renewcommand{\theequation}{6.\arabic{equation}}
\setcounter{equation}{0}

\section*{VI. A stronger  criterion for the quantum diffusion}
\noindent
The presence of sizeable quantum diffusion enters in  another relevant consideration and provides a stronger criterion. 
Assume a Gaussian form for the PBH mass function 
\be
\label{ggg1}
\beta_{\rm prim} (M)\simeq  \frac{\sigma_{\cal R}}{\sqrt{2\pi}{\cal R}_c}e^{-{\cal R}_c^2/2\sigma_{\cal R}^2}.
\ee
Suppose one fine-tunes the parameters of the inflaton potential to produce the right amount of PBH as dark matter, but  without accounting for the quantum diffusion and therefore the spread of the inflaton velocities.  

Practitioners of the production of PBHs as dark matter  in single-field models of inflation know that a considerable
fine-tuning is needed in any model to produce the right amount of dark matter in the form of PBHs. Any deviation from the fine-tuned
set of parameters due to the uncertainty caused by the quantum diffusion  will lead to huge variations of the PBH primordial mass fraction (as well as the ignorance on the non-Gaussian corrections do).
Let us take therefore into account the spread now on the PBH mass fraction itself. 

\subsection*{Linear potential}
\noindent
Assuming that   $\sigma_{{\cal R}}\simeq {\cal P}^{1/2}_{{\cal R}_{\rm pk}} \sim
(H/2\pi\Pi_\star)$,  the PBH mass fraction has an average induced by quantum diffusion equal to 
\be
\langle\beta_{\rm prim} (M)\rangle=\int\,\dd\Pi P_{\Pi}(\Pi)\left.\beta_{\rm prim} (M)\right|_{\rm Gaussian}.
\ee
Using a Gaussian distribution for $\Pi_\star$ with spread $\langle(\Delta\Pi)^2\rangle^{1/2}$, we get
\be
\langle\beta_{\rm prim} (M)\rangle=
\frac{H^3\sigma_{\cal R}}{\sqrt{2\pi}\left(H^2+4\pi^2{\cal R}_c^2\langle(\Delta\Pi)^2\rangle\right)^{3/2}{\cal R}_c}e^{-\frac{H^2{\cal R}_c^2}{2\left(H^2+4\pi^2{\cal R}_c^2\langle(\Delta\Pi)^2\rangle\right)\sigma_{\cal R}^2}}.
\ee
Notice that the average value of the PBH primordial abundance gets shifted with respect to the  expression (\ref{ggg1}) precisely because the
distribution of the inflaton velocity has a nonvanishing width.

We may define  a fine-tuning parameter $\Delta_{\rm qd}$ defined through the ratio of the averaged mass fraction in the presence of diffusion and the mass fraction in the absence of diffusion as
\be\label{secondcrit}
\frac{\langle\beta_{\rm prim} (M)\rangle}{\beta_{\rm prim}(M)}=e^{\Delta_{\rm qd}}.
\ee
Essentially, this fine-tuning parameter says how far is the average of the distribution of $\beta_{\rm prim} (M)$ from the classical value computed in the absence of
quantum diffusion.
We find that
\be
\Delta_{\rm qd}\simeq- \frac{{\cal R}_c^2}{2\sigma_{\cal R}^2}\left(\frac{\varepsilon}{1+\varepsilon}\right),\,\,\,\,\varepsilon=-\frac{4\pi^2{\cal R}_c^2\langle(\Delta\Pi)^2\rangle}{H^2}.
\ee
Imposing that the calculation is done in the absence of diffusion is trustable requires $|\Delta_{\rm qd}|\lsim 1$, or
\be
\label{2c}
\boxed{
\frac{\langle(\Delta\Pi)^2\rangle^{1/2}}{H}\lsim \frac{\sigma_{\cal R}}{\sqrt{2}\pi{\cal R}_c^2}\simeq 10^{-2}\left(\frac{\sigma_{\cal R}}{0.1}\right)\left(\frac{1.3}{{\cal R}_c}\right)^2.}
\ee
For a linear potential this bound is violated by the fact that the spread in the velocity is  $\sqrt{D/6}/H\simeq 0.2$.
One might think to reduce the fine-tuning by,  for instance, decrease the value of ${\cal R}_c$, however one should also recall that in order to get the right
amount of dark matter in the form of PBH, $\beta_{\rm prim}\simeq  10^{-16}$,  one needs $\sigma_{\cal R}/{\cal R}_c\sim 1/8$ and therefore decreasing  ${\cal R}_c$
leads to a strong decrease in $\sigma_{\cal R}$. Alternatively, one can fix the spread in the velocity to be $\sqrt{D/6}/H\simeq 0.2$ and, imposing $|\Delta_{\rm qd}|\lsim 1$, find a lower  bound on the
square root of the variance
\be
\sigma_{\cal R}\gsim \frac{2}{\sqrt{3}}{\cal R}_c^2\simeq 2\left(\frac{{\cal R}_c}{1.3}\right)^2,
\ee
which signals  the difficulty of avoiding the impact of the quantum noise.

%
%
%
%
%

\subsection*{Non-attractor:  more physical cases}
\noindent
For a more realistic potential, like the one in Ref. \cite{sone3}, we have seen that the spread in the velocities  at the end of the non-attractor phase
is as large as $\langle(\Delta\Pi)^2\rangle^{1/2}/H\simeq 0.4$. To assess the impact of the quantum diffusion on the PBH abundance, we have proceeded as follows. We have set the parameters of the model as in section II, see Tab. \ref{tab1},  in such a way to reproduce the right  abundance 
for the PBHs to be dark matter and for the potential to be consistent with the CMB constraints on the power spectrum at the reference scale of $k_{\rm CMB}= 0.05\, {\rm Mpc}^{-1}$ (the spectral index and the tensor to scalar ratio computed in the slow-roll region are as well in agreement with current data).

 The PBH abundance has been calculated using the  density contrast $\Delta(\vec x)=(4/9a^2H^2)\nabla^2\zeta(\vec x)$ with threshold    $\Delta_c\simeq 0.45$ \cite{bb} where the variance is defined as 
\be
\sigma_\Delta ^2(R_{H})= \frac{16}{81} \int _0^\infty \dd  \ln q  (qR_H)^ 4  W^2(qR_H){\cal P}_{\cal R}(q),
\ee
where  $W(qR_H)$ is a Gaussian window function smoothing out the density contrast on the comoving horizon length
 $R_H=1/aH$.  The Gaussian approximation  of the primordial mass fraction
\be
\label{ggg}
\beta_{\rm prim} (M)\simeq  \frac{\sigma_\Delta}{\sqrt{2\pi}{\Delta_c}}e^{-\Delta_c^2/2\sigma_{\Delta}^2},
\ee
gives  $\beta_{\rm prim}(10^{-15} M_{\odot}) \simeq 3 \cdot10^ {-16}$ and therefore the right dark matter abundance.
We  have then   included the quantum diffusion, run $10^4$ realisations of the stochastic background evolution and for each of them we have calculated the
primordial PBH abundance $\beta^{\rm qd}_{\rm prim}$. 

\begin{figure}[ht!]
 \centering
    \begin{minipage}{0.49\textwidth}
      \includegraphics[scale=0.9]{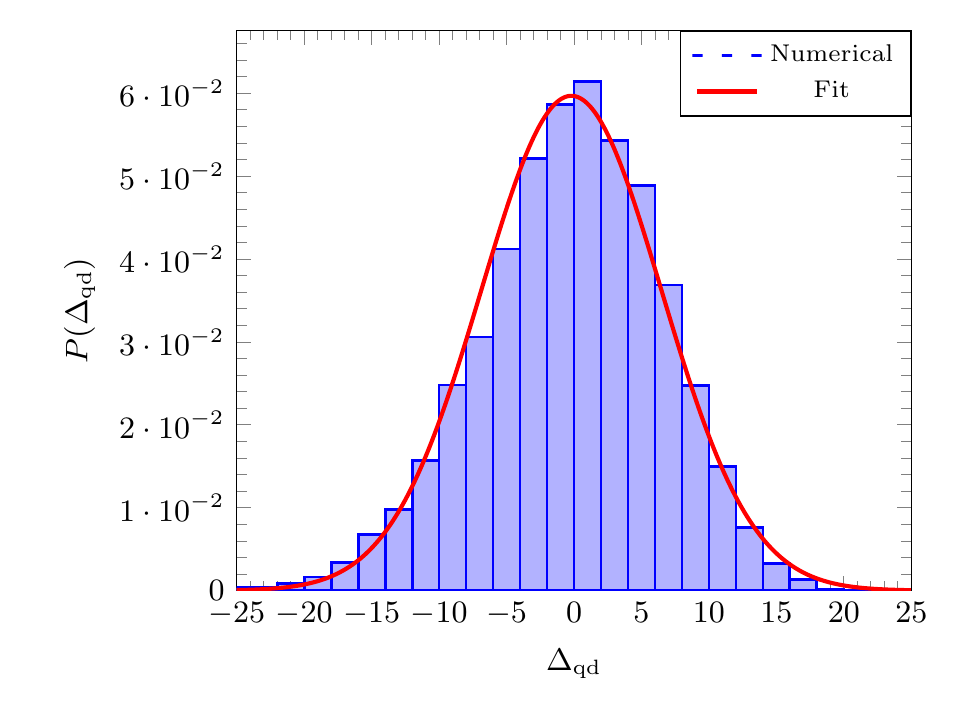}
      \end{minipage}
          \begin{minipage}{0.49\textwidth}
      \includegraphics[scale=0.9]{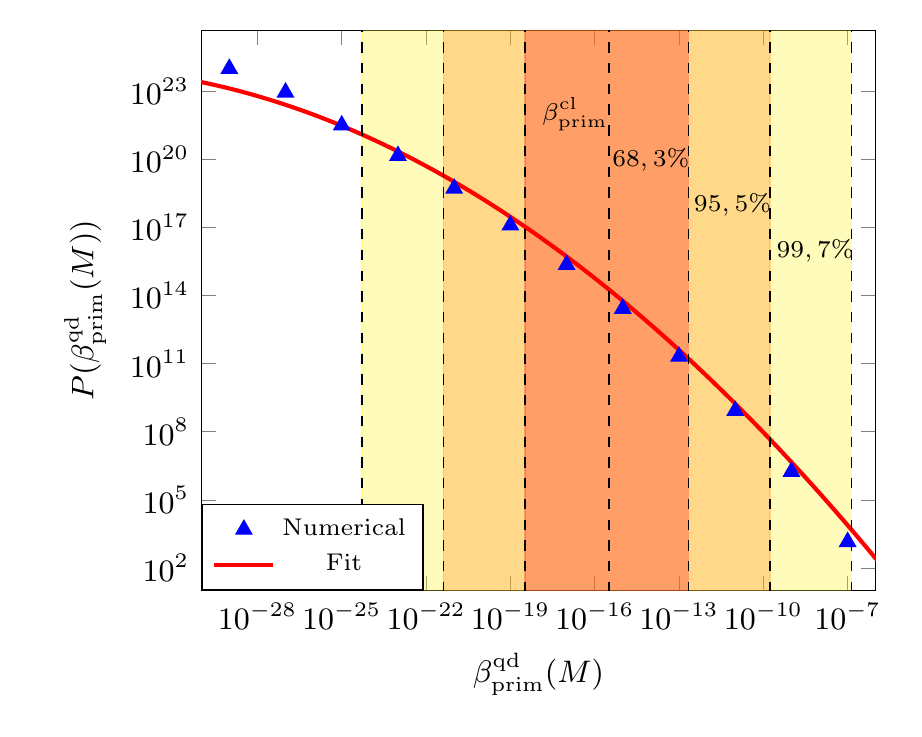}
      \end{minipage}
     \caption{\small The probability density of $\Delta_{\rm qd}$, which is nearly Gaussian distributed around the  classical value determined ignoring quantum diffusion,  and of  $\beta^{\rm qd}_{\rm prim}$ for model in Ref. \cite{sone3}. The results are derived from $ 10^4$ realisations of the stochastic evolution.}      \label{betac}
\end{figure}

Our results show  that $\ln\beta^{\rm qd}_{\rm prim}$ is approximately Gaussian distributed around the value of $\beta^{\rm cl}_{\rm prim}$ computed using the classical inflaton evolution, and with a standard deviation $\sigma_{\beta_{\rm prim}^{\rm qd}}$, see Fig. \ref{betac}.  This is only an approximation because there is a small skewness shifting the
average slightly away from its classical value. This means that $\beta^{\rm qd}_{\rm prim}$ is nearly distributed as a log-normal distribution. Extending what we have done previously, we can introduce a fine-tuning parameter defined to be
\be
\Delta_{\rm qd}=\ln\frac{\beta^{\rm qd}_{\rm prim} }{\beta^{\rm cl}_{\rm prim}}.
\ee
%
%
%
%
\begin{figure}[ht!]
 \centering
    \begin{minipage}{0.49\textwidth}
      \includegraphics[scale=0.9]{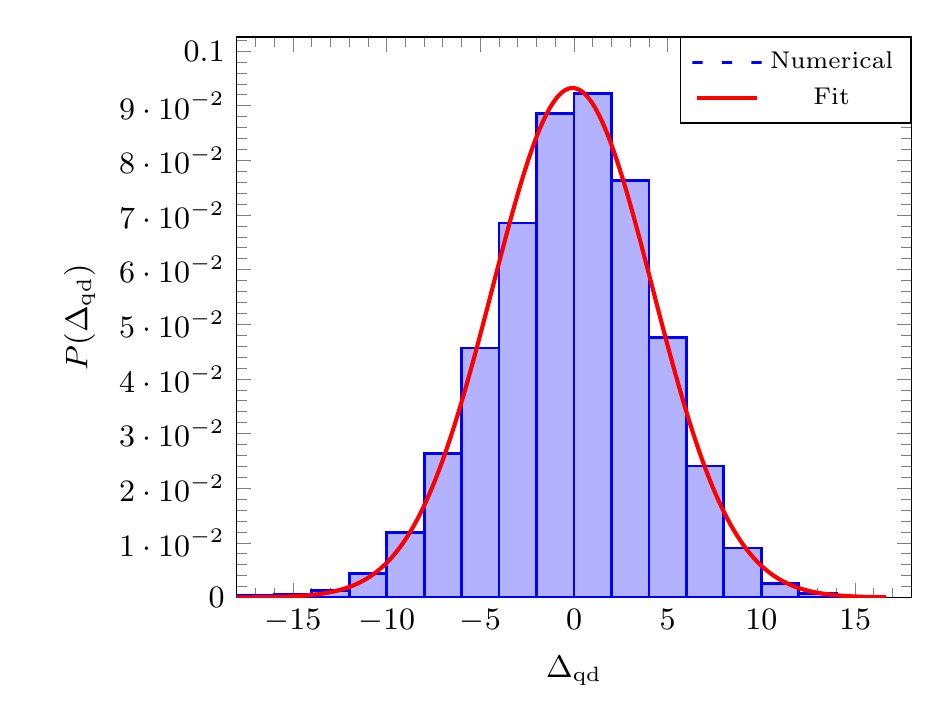}
      \end{minipage}
          \begin{minipage}{0.49\textwidth}
      \includegraphics[scale=0.9]{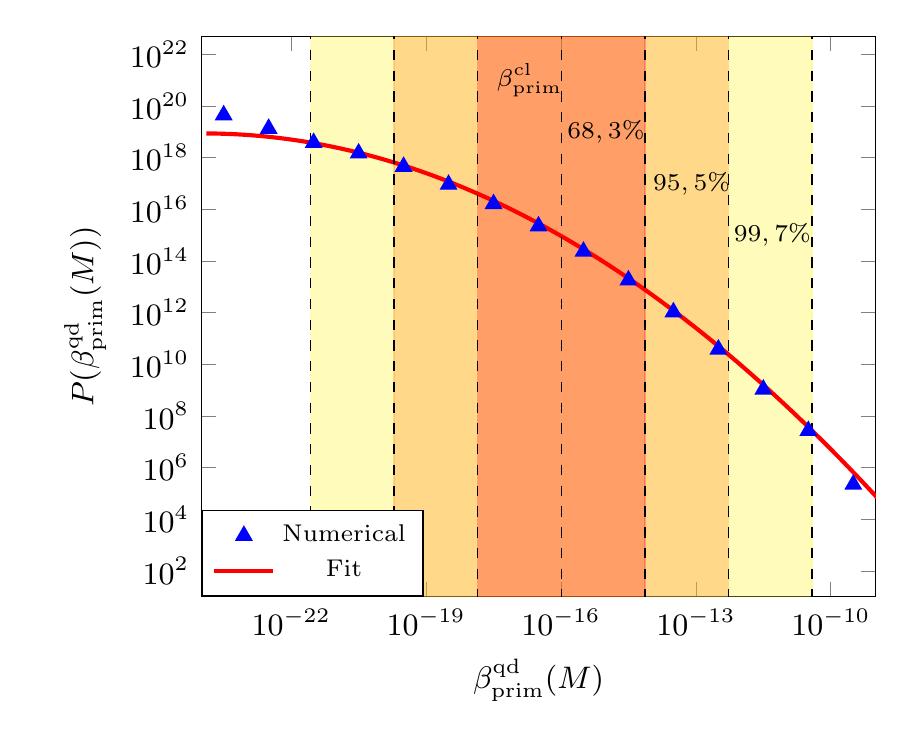}
      \end{minipage}
     \caption{\small The probability density of $\Delta_{\rm qd}$, which is nearly Gaussian distributed around the  classical value determined ignoring quantum diffusion,  and of  $\beta^{\rm qd}_{\rm prim}$ for model in Ref. \cite{sone4}. The results are derived from $ 10^4$ realisations of the stochastic evolution.}      \label{betat}
\end{figure}
This quantity is distributed like a Gaussian and is a measure of how close the distribution of the  PBH mass fraction is peaked around the classical value. Therefore
$\Delta_{\rm qd}$ is (nearly) centered around zero and within $p\sigma_{\beta^{\rm qd}_{\rm prim}}$ it acquires values

 \be\label{df}
  -p\sigma_{\beta^{\rm qd}_{\rm prim}}\lsim \Delta_{\rm qd}(p)\lsim   p\sigma_{\beta^{\rm qd}_{\rm prim}}.
  \ee
  The  values of $\Delta_{\rm qd}(p)$ are summarised in Tab. \ref{tabdelta}. Notice that the range is not totally symmetric because of the small skewness.
\begingroup
\setlength{\tabcolsep}{10pt} 
\begin{table}[ht!]
\footnotesize
\begin{center}
\caption{{\small Detailed values of the $\Delta_{\rm qd}(p)$ as defined in Eq. \eqref{df} and their corresponding values of $\beta_{\rm prim}$ for models \cite{sone3,sone4}. }}\label{tabdelta}
\begin{tabular}{|c|c|c|c|c|c|c|}
\hline
{\bf Model} \cite{sone3}&$p= 1$&$p=- 1$&$p= 2$&$p= -2$&$ p= 3$&$p= -3$\\
\hline
$\Delta_{\rm qd}(p) $&$-6.49 $&$ 6.88$&$ -13.17$& $13.56 $&$-19.85$&$20.25$\\
\hline
$\beta_{\rm prim} ^{\rm qd} (p)$&$2.13\cdot10^{-13}$&$3.33\cdot10^{-19}$&$1.70\cdot10^{-10}$&$4.16\cdot10^{-22}$&$1.36\cdot10^{-7}$&$5.20\cdot10^{-25}$\\
\hline
\hline
{\bf Model} \cite{sone4}&$p= 1$&$p=- 1$&$p= 2$&$p= -2$&$ p= 3$&$p= -3$\\
\hline
$\Delta_{\rm qd}(p) $&$-4.20 $&$ 4.36$&$ -8.48$& $8.64 $&$-12.76$&$12.92$\\
\hline
$\beta_{\rm prim} ^{\rm qd} (p)$&$7.33\cdot10^{-15}$&$1.41\cdot10^{-18}$&$5.29\cdot10^{-13}$&$1.95\cdot10^{-20}$&$3.81\cdot10^{-11}$&$2.70\cdot10^{-22}$\\
\hline
\end{tabular}
\end{center}
\end{table}%
\endgroup
\noindent
We observe that the criterion $|\Delta_{\rm qd}|\lsim 1$ is grossly violated and the value of $\beta^{\rm qd}_{\rm prim} (M)$ violently deviates from the classical value due to the effect of quantum diffusion on the evolution of the background.
In other words the values of the PBH mass fraction violently fluctuate around an average which is very different from the classical value thought to be needed to get  the right abundance
of the dark matter in the form of PBH. 

Similar results are obtained for the model in Ref. \cite{sone4}, they are shown in Fig. \ref{betat}. For the sake of comparison we have  used the same reference values as in Ref. \cite{sone4}.  We notice that the dispersion of $\beta_{\rm prim} (M)$ is less prominent. However, this is only due to the fact that in Ref. \cite{sone4}  a smaller threshold, $\Delta_{c} \simeq 0.3$, has been adopted, leading to smaller values of the variances to reproduce the right amount of dark matter in the form of PBHs. As a consequence, the  impact of quantum diffusion is relatively smaller. Still,  the criterion is violated as we can see  from Tab. \ref{tabdelta}. We also remark that higher values of $\Delta_c=(0.4-0.7)$ \cite{bb} are used  in the literature and therefore even larger values of  $\Delta_{\rm qd}$ will be obtained.

 Our results make us confident that, while in principle   conclusions   might depend on the exact
values of the square root of the variance  $\sigma_{\Delta}$ and threshold ${\Delta}_c$, the corresponding 
 $|\Delta_{\rm qd}|$ will in general be too large. This is because changing the parameters of the model to get  new variances with some  new thresholds does not reduce significantly the spread of $\ln\beta_{\rm prim}$. 
Therefore, while our  results are specific of the models we have considered, we believe the conclusions apply to any model where the inflaton field crosses a plateau with an inflection point in order to generate a spike in the power spectrum and give rise to PBHs.
%

We expect therefore that   the standard (classical) picture to evaluate the dark matter abundance in terms of PBHs is  significantly  altered.

\renewcommand{\theequation}{7.\arabic{equation}}
\setcounter{equation}{0}

\section*{VII. Conclusions}
 There is a lot of interest in the cosmology community for the possibility that the dark matter is formed by PBHs. Their origin
 might be ascribed to the same mechanism giving rise to the CMB anisotropies and large-scale scale structure, i.e. a period of inflationary accelerated expansion during the
 early stages of evolution of the universe. In single-field models the power spectrum of the curvature perturbation might increase at small scales if the inflaton crosses a region
 which is flat enough and various models in the literature have been proposed recently. 
 
  In this paper we have discussed the role of quantum diffusion in the determination of the final abundance of PBHs. Quantum diffusion necessarily acquires importance when the
  force induced by the inflaton potential becomes tiny during the dynamics of the inflaton field. We have analysed both analytically and numerically
  the impact of diffusion and concluded that in realistic models it can significantly affect the capability of making a firm prediction of the PBH abundance. This is because
  the velocity of the inflaton field turns out to be distributed  around its classical value with a  spread which has an exponential impact on the PBH mass fraction.
  
  While by itself the mass fraction does not say anything about the spatial distribution of the PBHs, we expect  that different regions of the universe upon PBH formation would be populated with different relative abundances, thus changing the prediction for how much dark matter there is or its subsequent evolution.

\bigskip
\bigskip
\noindent
\subsection*{Acknowledgments}
\noindent
We  thank M. Cicoli, F. G. Pedro, and G. Tasinato  for many interactions about their work (Refs. \cite{sone3} and \cite{sone4} respectively) and feedback on our draft. We also thank  A. Linde for interactions about the PBH formation probability. We also thank C. Germani for disscussions.
  A.R. is  supported by the Swiss National Science Foundation (SNSF), project {\sl Investigating the Nature of Dark Matter}, project number: 200020-159223. A.K. thanks the Cosmology group at the D\'epartement de Physique 
  Th\'eorique  at the  Universit\'e de Gen\`eve for the kind hospitality and financial support. 
%
%

\section*{Appendix A: the curvature perturbation, the Schwarzian derivative and the dual transformation}
\setcounter{equation}{0}
\renewcommand{\theequation}{A.\arabic{equation}}
\noindent
In this Appendix we elaborate further the issue of why the power spectrum during the non-attractor phase is indeed flat. This Appendix does not contain
some new material with respect to the literature.

Our starting point is  the equation for the curvature perturbation on comoving
hypersurfaces  ${\cal R}$
\be
\label{ff}
{\cal R}''+2\frac{z'}{z}{\cal R}'+k^2{\cal R}=0,
\ee
where for convenience the prime denotes in this Appendix  and in the following one the conformal time derivative ${\rm d}/{\rm d}\tau$ and $z=a\dot\phi/H$ (the dot  denotes the cosmic  time derivative).  The function $z$ satisfies the following equation 
\be
\label{zs}
\frac{z''}{z}=2a^2H^2\left(1+\epsilon-\frac{3}{2}\eta+\epsilon^2-2\epsilon\eta+\frac{1}{2}\eta^2+\frac{1}{2}\xi^2\right)=2a^2H^2\left(1+\frac{5}{2}\epsilon+\epsilon^2-2\epsilon\eta-\frac{1}{2}\frac{V_{,\phi\phi}}{H^2}\right),
\ee
where
\begin{eqnarray}
\label{sr}
\epsilon&=&-\frac{\dot H}{H^2},\nonumber\\
\eta&=&-\frac{\ddot \phi}{H\dot\phi},\nonumber\\
\xi^2&=&3(\epsilon+\eta)-\eta^2-\frac{V_{,\phi\phi}}{H^2}.
\end{eqnarray}
As long as  slow-roll is attained, 
one can make use of the corresponding slow-roll parameters deduced from the form of the potential
\begin{eqnarray}
\epsilon_V&=&\frac{1}{2}\left(\frac{V_{,\phi}}{V}\right)^2,\nonumber\\
\eta_V&=&\frac{1}{3}\frac{V_{,\phi\phi}}{H^2},\nonumber\\
\xi^2_V&=&3\epsilon_V-\eta_V^2,
\end{eqnarray}
where  
 the dynamics around Hubble crossing is   dominated by the exponentially growing friction term proportional to  ${\cal R}'$, and the solution to Eq. (\ref{ff})   is  well approximated by 
\be
{\cal R}(\tau)={\rm constant}\,\,\,\,{\rm and}\,\,\,\, \frac{{\cal R}'(\tau)}{aH}\sim \left(\frac{\tau}{\tau_k}\right)^2,
\ee
where 
$\tau_k$ indicates the  value of the conformal time at which the  comoving wavelength $\sim 1/k$  leaves the comoving Hubble radius. 

In the case in which one is interested in the generation of PBH from sizeable curvature fluctuations at small scales, a violent departure from the slow-roll must occur.  In particular, if after Hubble crossing the friction term proportional to $z'/z$  changes its sign from positive to negative, it may become  a  driving term.  This can have significant  effects on modes which leave or have  left the Hubble radius during this transient and non-attractor  epoch and thus induce a growth of the curvature perturbations \cite{ll,leachetal}. A necessary, but not sufficient condition, to have PBH generation from single-field models is therefore  the presence of a transient period for which
\be
\frac{z'}{z}=aH\left(1+\epsilon-\eta\right)<0.
\ee
During such stage the  function $z$ reaches a local extremum (a maximum or a minimum depending upon the sign of $\dot\phi$) at some time  whenever 
\be
1+\epsilon-\eta=0,
\ee
Since $\epsilon$ is always positive, the presence of a transient stage  implies that $\eta$ must be at least unity, signalling a breakdown of the slow-roll conditions.

In order to simplify the problem of dealing with a non-attractor phase necessary to generate a large amount of PBHs, we start by  noticing 
that, upon the redefinition 
\be
{\cal R}=\frac{\widetilde z}{z}\widetilde{\cal R},
\ee
the quantity  $\widetilde{\cal R}$ satisfies the same equation of ${\cal R}$
\be
\widetilde{\cal R}''+2\frac{\widetilde{z}'}{\widetilde z}\widetilde{\cal R}'+k^2\widetilde{\cal R}=0,
\ee
as long as
\be
\label{rel}
\frac{z''}{z}=\frac{\widetilde{z}''}{\widetilde z}.
\ee
The  transformation from $z$ to $\widetilde{z}$ which satisfies the relation  (\ref{rel}) has been nicely worked out in Ref. \cite{wands} and called a dual  transformation. It reads
\be
\label{fun}
\widetilde{z}(\tau)=C_1z(\tau)+C_2 z(\tau)\int^\tau\frac{{\rm d}\tau'}{z^2(\tau')}.
\ee
In fact this transformation is a  property inferred from the so-called Schwarzian derivative \cite{sch}, which we briefly summarise in the next subsection

\subsection*{The Schwarzian derivative}
\noindent
Given a function $f(\tau)$, the 
Schwarzian derivative  is defined as 
\begin{eqnarray}
f\mapsto S[f]=\frac{f'''}{f'}-\frac{3}{2}\left(\frac{f''}{f'}\right)^2. 
\end{eqnarray}
A property of the Schwarzian is that it is invariant under the transformation
\begin{eqnarray}
 \widetilde{f}=\frac{a\, f+b}{c\, f+d}, ~~~~~~ad-b c\neq 0, \label{tr}
\end{eqnarray}
that is 
\begin{eqnarray}
S[\widetilde{f}]=S[f]. \label{sl}
\end{eqnarray}
Note that the symmetry (\ref{sl}) is just SL$(2,\mathbb{R})$ up to a rescaling of $f$.
Consider now a differential equation 
\begin{eqnarray}
u''+q(\tau)u=0.  \label{de}
\end{eqnarray}
It can easily be seen that 
\begin{eqnarray}
q(\tau)=\frac{1}{2}S[f],  \label{q1}
\end{eqnarray}
where 
\begin{eqnarray}
f(\tau)=\int^\tau \frac{{\rm d} \tau'}{u^2(\tau')}.\label{fu}
\end{eqnarray}
Indeed, from Eq. (\ref{fu}) we find that 
\begin{eqnarray}
u=\frac{1}{\sqrt{f'}}, \label{uf}
\end{eqnarray}
and therefore,  
\begin{eqnarray}
u''=-\frac{1}{2} S[f] u,
\end{eqnarray}
which is nothing else than Eq.(\ref{de}) and Eq. (\ref{q1}). 
Now, 
since the Schwarzian is invariant under the transformation (\ref{tr}), we have that 
\begin{eqnarray}
\frac{u''}{u}=-\frac{1}{2}S[f]=-\frac{1}{2}S[\widetilde{f}]=\frac{\widetilde{u}''}{\widetilde u},
\end{eqnarray}
where 
\begin{eqnarray}
\widetilde{u}= \frac{1}{\sqrt{\widetilde{f}'}}.  \label{ut}
\end{eqnarray}
Then, using Eqs. (\ref{tr}), (\ref{uf}) and (\ref{ut}) we find that 
\begin{eqnarray}
\widetilde{u}= u (C1+C_2 f),~~~~C_1=\frac{d}{\sqrt{ad-bc}}, ~~~C_2=\frac{c}{\sqrt{ad-bc}},  
\end{eqnarray}
which, by using Eq.(\ref{fu}), is written as
\begin{eqnarray}
\widetilde{u}(\tau)=C_1u(\tau)+C_2u(\tau) \int^{\tau}\frac{{\rm d}\tau'}{u^2(\tau)},
\end{eqnarray}
which is nothing else than the dual transformation found in Ref. \cite{wands}.

Going back to the transformation (\ref{fun}), 
%
the power spectrum of the comoving curvature perturbation at the end of inflation reads
\be
\boxed{
{\cal P}_{\cal R}\Big|_{\textrm{end of inflation}}=\frac{\widetilde{z}}{z}{\cal P}_{\widetilde{\cal R}}\Big|_{\textrm{end of inflation}},}
\ee
from which we deduce that the power spectrum of the comoving curvature perturbation ${\cal R}$ is flat, a property that is inherited by the power spectrum of  $\widetilde{{\cal R}}$ for which the dynamics is of the slow-roll nature. The question now  is  what is the most suitable dual transformation  to perform in order to simplify the computation of the power spectrum for those modes which exit the Hubble radius during the non-attractor phase and which are ultimately responsible for the production of  PBHs when these curvature perturbations re-enter the Hubble radius during the radiation phase.

As we have mentioned already several times, the production of   PBHs may  originated from the enhancement of the curvature power spectrum  below a certain length scale. This can be achieved by a temporary abandonment of the slow-roll condition.
When the inflaton field follows slow-roll and $\dot\phi$ is approximately constant, the function $z=a\dot\phi/H$ grows 
%
\be
\label{sr1}
z\sim \frac{1}{\tau}
\,\,\,\, \textrm{during slow-roll}.
\ee
During the non-attractor phase, when the inflaton field experiences an approximately  flat potential and $V_{,\phi}$ can be neglected, it satisfies the equation of motion
\be
\phi''+2 {\cal H}\phi'\simeq 0\,\,\,\, \,\,\,({\cal H}=aH),
\ee
and consequently $\phi'\sim \tau^2$, or
\be
z=\frac{a\dot\phi}{H}=\frac{\phi'}{H}\sim \tau^2
\,\,\,\,\textrm{during the non-attractor phase}.
\ee
It is this rapid fall of $z$ which allows the possibility of enhancing the power spectrum. When the non-attractor phase is over, the slow-roll conditions
are attained again and one recovers the behaviour in Eq. (\ref{sr1}). In terms of the friction term $z'/z$ one has 
\be
\frac{z'}{z}\simeq aH\left\{ 
\begin{array}{ll} 
1 \ \ &\textrm{during slow-roll}, \\ 
-2 \ \ & \textrm{during the non-attractor phase}.
\end{array} 
\right.
\ee
Let us now use the dual transformation (\ref{fun}) where we choose the lower limit of the integral to be  $\tau_0$, the initial conformal time for the non-attractor phase. In such a case, we find  
%
%
%
%
%
%
%
\be
\frac{\widetilde{z}'}{\widetilde z}=\frac{z'}{z}+\frac{C_2}{z}\frac{1}{C_1z +C_2 z\int^\tau_{\tau_0}\,{\rm d}\tau'/z^2(\tau')}.
\ee
We can now choose $C_1=1$ and  compute this expression during the non-attractor phase for those modes which enter the Hubble radius during the non-attractor phase
\be
\frac{1}{aH}\frac{\widetilde{z}'}{ \widetilde z}=-2+\frac{C_2}{aH z}\cdot\frac{1}{z +C_2 z\int^\tau_{\tau_0}\,{\rm d}\tau'/z^2(\tau')}.
\ee
Setting $a=a_0(\tau_0/\tau)$ and $z(\tau)=z_0 (\tau/\tau_0)^2$,  we obtain
\be
\frac{1}{aH}\frac{\widetilde{z}'}{ \widetilde z}=-2+\frac{C_2}{a_0 z_0H}\left(\frac{\tau_0}{\tau}\right)\frac{1}{z_0 (\tau/\tau_0)^2 -(C_2/3z_0) (\tau_0^2/\tau)}.
\ee
where in the last passage we have neglected the subleading  term  $\sim 1/\tau^3_0$. 
Taking $-\tau<-\tau_0$ (recall that $\tau<0$) and recalling that $a_0=-1/(H\tau_0)$, one finally obtains
\be
\frac{1}{aH}\frac{\widetilde{z}'}{ \widetilde z}\simeq-2+3=1\,\,\,\, \textrm{(during the non-attractor phase)}.
\ee
This demonstrates that the choice
\be
\label{fun3}
\widetilde{z}(\tau)=z(\tau)+C_2 z(\tau)\int^\tau_{\tau_0}\frac{{\rm d}\tau'}{z^2(\tau')},
\ee
maps the non-attractor phase into  a slow-roll phase for the curvature perturbation $\widetilde{{\cal R}}$ and one can conclude that the power spectrum of the curvature perturbation for those modes entering the Hubble radius during the non-attractor phase is dictated by a slow-roll dynamics and therefore is flat. Its  amplitude
is however magnified by a factor  $\widetilde{z}(\tau_{\rm e})/z(\tau_{\rm e})$.

To elaborate  further and find a useful prescription, let us consider, as we also did in the main text,  Starobinsky's model \cite{Starobinsky92} where the inflaton field reaches a non-attractor phase after a slow-roll era (and eventually enters afterwards another slow-roll phase).

\subsection*{Slow-roll phase before the non-attractor phase}
\noindent
If we indicate by $\phi_0$ the moment at which the first slow-roll phase ends and the non-attractor phase starts, we can Taylor expand the inflaton potential as  
\be
V(\phi)\simeq V_0\left(1+\sqrt{2\epsilon_{+}}(\phi-\phi_0)\right)+\cdots\,\,\,\,{\rm for}\,\,\,\,\phi>\phi_0,
\ee
where $\epsilon_{+}$ is the slow-roll parameter during the first slow-roll phase.  The corresponding parameter $z$ reads
\be
3{\cal H}\phi'=-V_{,\phi}a^2\,\,\,\,{\rm and}\,\,\,\,
z_+(\tau) \simeq -a_0\sqrt{2\epsilon_{+}}(\tau_0/\tau),
\ee
having indicated $\tau_0$ and $a_0$ is the conformal time and the scale factor when $\phi=\phi_0$, respectively.
\subsection*{Non-attractor phase}
\noindent
For $\phi<\phi_0$ the potential is Taylor expanded as 
\be
\label{lin}
V(\phi)\simeq V_0\left(1+\sqrt{2\epsilon_{-}}(\phi-\phi_0)\right)+\cdots \,\,\,\,{\rm for}\,\,\,\,\phi<\phi_0.
\ee
The dynamics leads to
%
\be
\frac{3{\cal H}\phi'}{V_0a^2}=-\sqrt{2\epsilon_{-}} -(\sqrt{2\epsilon_{+}}-\sqrt{2\epsilon_{-}})(\tau/\tau_0)^3
\ee
and
\be
\label{approxz} 
z_-(\tau) \simeq -a_0\left(\sqrt{2\epsilon_{-}}(\tau_0/\tau) + (\sqrt{2\epsilon_{+}}-\sqrt{2\epsilon_{-}})(\tau/\tau_0)^2\right). 
\ee
If $\epsilon_{+}\gg \epsilon_{-}$, there is a prolonged non-attractor phase where the second term in the above equation dominates over the first one.
It is easy to show that $z_-$ reaches a   maximum at the point 
\be
(\tau_0/\tau_{\rm m})^3=2\frac{\sqrt{2\epsilon_+} - \sqrt{2\epsilon_-}}{\sqrt{2\epsilon_-}}
\simeq 2\sqrt{\frac{\epsilon_+}{\epsilon_-}},
\ee
corresponding to   $z_-(\tau_{\rm m})\approx
(\epsilon_-/ \epsilon_+)^{1/3}z(\tau_0)$ and  causing  a sizeable
change in ${\cal R}$ on super-Hubble  scales if $z_-(\tau_{\rm m})$ is very tiny. Notice that the smallness of $\epsilon_-$ parametrises the duration of the non-attractor phase
from $\tau_0$ to $\tau_\star$.
Let us now consider the duality transformation (\ref{fun}) with again lower limit $\tau_0$ in the integral. 
%
We deduce
\begin{eqnarray}
\widetilde{z}_-(\tau)=-C_1 a_0 \sqrt{2\epsilon_-} (\tau_0/\tau) +\frac{a_0^3 \sqrt{2\epsilon_-}(\sqrt{2\epsilon_-}-\sqrt{2\epsilon_+})C_1+3 C_2H^3}{a_0^2\sqrt{2\epsilon_-} }(\tau/\tau_0)^2. \label{zexp}
\end{eqnarray}
We are free to choose $C_2$ such that the second term on the right-hand side  of Eq. (\ref{zexp}) vanishes, which happens for
\begin{eqnarray}
C_2=\frac{a_0^3\sqrt{2\epsilon_-}(\sqrt{2\epsilon_+}-\sqrt{2\epsilon_-})}{3H^3} \, C_1, 
\end{eqnarray}
and hence
\begin{eqnarray}
\widetilde{z}_-(\tau)=-C_1 a_0 \sqrt{2\epsilon_-} (\tau_0/\tau). \label{zexp1}
\end{eqnarray}
We are also free to make the dual transformation only for $\phi<\phi_0$ and therefore we match $z_+$ with the new $\widetilde{z}_-$ at $\tau_0$ and  find
\begin{eqnarray}
C_1=\sqrt{\frac{\epsilon_+}{\epsilon_-}}. \label{c1}
\end{eqnarray}
Therefore we have a single slow-roll parameter
\begin{eqnarray}
z=z_+=\widetilde{z}_-= -a_0\sqrt{2\epsilon_+}(\tau_0/\tau),  
\end{eqnarray}
throughout all the  evolution.
This implies that the power spectrum for the $\widetilde{\cal R}$ not only remains  constant after Hubble crossing, but also can be computed using the slow-roll approximations and it reads  
\begin{eqnarray}
{\cal P}_{\widetilde{\cal R}}^{1/2}= \frac{3H^3}{2\pi V_0\sqrt{2\epsilon_+}}\Big|_{k=aH},
\end{eqnarray}
even during the  non-attractor phase. The power spectrum therefore evolves as
\be
\label{evolv}
{\cal P}^{1/2}_{{\cal R}}(\tau)=\frac{\widetilde{z}(\tau)}{z(\tau)}{\cal P}^{1/2}_{\widetilde{{\cal R}}}=
\frac{\widetilde{z}_-(\tau)}{z_-(\tau)}{\cal P}^{1/2}_{\widetilde{{\cal R}}}=
\frac{1}{\sqrt{2\epsilon_-/2\epsilon_+}+(1-\sqrt{2\epsilon_-/2\epsilon_+})(\tau/\tau_0)^3}{\cal P}^{1/2}_{\widetilde{{\cal R}}},
\ee
Defining by $\tau_\star$ the end of the non-attractor phase and computing the power spectrum just after $\tau_\star$ (recall that the conformal time is negative and therefore
$-\tau_\star\ll -\tau_0$))
we find that immediately after  the non-attractor phase
 \begin{eqnarray}
 {\cal P}^{1/2}_{{\cal R}}(\tau\gsim\tau_\star)=\frac{\sqrt{2\epsilon_+}}{\sqrt{2\epsilon_-}}{\cal P}^{1/2}_{\widetilde{{\cal R}}}= \frac{3H^3}{2\pi V_0\sqrt{2\epsilon_-}}. 
 \end{eqnarray}
 At the end of the non-attractor phase  therefore one finds  
  \begin{eqnarray}
 \label{per}
 \boxed{
{\cal P}^{1/2}_{{\cal R}}(\tau\gsim\tau_\star)=\left(\frac{H}{2\pi\Pi_\star}\right)=\left.\frac{3H^3}{2\pi V_{,\phi}}\right|_{k=aH},}
 \end{eqnarray}
 which provides the prescription to compute the power spectrum of the curvature perturbations for those modes crossing the Hubble radius deep in the  non-attractor phase. By tuning the slope of the potential
 one can in principle obtain a large enhancement of the power spectrum. 
 
 A few comments are in order. The last passage in Eq. (\ref{per}) is valid only if the subsequent
 slow-roll phase starts when the velocity of the inflaton field has already settled to its slow-roll value proportional to $\sqrt{2\epsilon_-}$.
 We remind the reader   that the power spectrum does not further evolve during the subsequent transition between the non-attractor phase and the second slow-roll phase \cite{sasaki}.  The  prescription (\ref{per}) was already proposed in Ref. \cite{seto} (see also Refs. \cite{wands,leachetal,ll})  to deal with the  singular case in which $\dot\phi=0$. In this sense the  results of this long Appendix  are not new, but  we have  given  an alternative and maybe more intuitive derivation. Furthermore, the 
  prescription (\ref{per}) can be used   for those modes which exit the Hubble radius deep in  the non-attractor phase and predicts a flat power spectrum  as the
dual $\widetilde{R}$ experiences a slow-roll dynamics. Said in other words, the power spectrum  must be flat since $\widetilde{z}''/\widetilde z=z''/z\simeq 2a^2H^2$ up to small correction ${\cal O}(\epsilon_V)$. If one wishes to compute the abundance of PBHs using single-field models where a non-attractor phase is necessary, the corresponding   power spectrum of the comoving curvature perturbation can be computed by simply evaluating it at Hubble crossing, even during the non-attractor phase, as long as
  one makes use of the slow-roll relation $\dot\phi=-V_{,\phi}/3H$; one can then account for the modes leaving the  Hubble radius when $\eta$ grows fast from tiny values to 3 using
  Eq. (\ref{peak}).
Finally, the  prescription is based on the fact that the non-attractor phase is long enough for the dynamics to established. If the plateau is short in field space, the inflaton
  field may arrive at it with an  excessive kinetic energy and roll away of it in a Hubble time or so.

\section*{Appendix B: from the non-attractor back to the slow-roll phase}
\setcounter{equation}{0}
\renewcommand{\theequation}{B.\arabic{equation}}
\noindent
The modes which have crossed the Hubble radius during the non-attractor phase are on super-Hubble scales during the eventual subsequent transition to a slow-roll phase with larger  slope in the potential. To see what happens to these modes
we follow  Ref. \cite{sasaki} and  model again the potential during the transition as 
\be
V(\phi)\simeq V_0\left(1+\sqrt{2\epsilon_\star}(\phi-\phi_\star)\right)+\cdots,
\ee
where we have  defined  $\phi_\star$ the field value at the end of the attractor phase. 
The equation of motion for the inflaton field during the transition epoch reads
\be
\phi''+3{\cal H}\phi'+3a^2\sqrt{2\epsilon_\star}=0,
\ee
whose solution  for initial velocity $\Pi_\star$ leads to
\be
z(\tau)
=-\frac{\Pi_\star}{18}\left(3(6+h)(\tau/\tau_\star)^2-3h\frac{\tau_\star}{\tau}\right),\,\,\, h=6\sqrt{2\epsilon_\star}/\Pi_\star.
\ee
The solution for the super-Hubble scale comoving perturbation during the transient epoch reads
\be
{\cal R}(\tau)=C_1+C_2\int^\tau\frac{\dd\tau}{z^2(\tau')}= C_1+C_2\frac{12\tau_\star }{2(6+h)\epsilon_-(-h+(6+h)\tau^3)}.
\ee 
This solution needs to be matched now with the solution for $\tau<\tau_\star$ which (apart from the standard $(H/2\pi)(1/\sqrt{2k^3})$ scales like 
$(\tau_\star/\tau)^3/\Pi_\star$. Matching the perturbations and their derivatives at $\tau_\star$, one gets
\be
\left.{\cal R}\right|_{\textrm{end of inflation}}=\left(\frac{H}{2\pi\sqrt{2 k^3}}\right)\frac{6+h}{h\Pi_\star}.
\ee
We see that  if the non-attractor phase is followed by another slow-roll phase for which  $|h|\gg 1$,  the curvature perturbations associated to the modes which are on super-Hubble scales during the transition will keep be enhanced as   $1/\sqrt{2\epsilon_-}$ and the prescription (\ref{per}) remains valid for those modes exiting the Hubble radius deep in the non-attractor phase \cite{sasaki}.

\section*{Appendix C: the role of non-Gaussianities}
\setcounter{equation}{0}
\renewcommand{\theequation}{C.\arabic{equation}}
\noindent
As we mentioned in the introduction, PBHs are born as large, but rare fluctuations of the curvature perturbation. As such, their abundance is extremely
sensitive to the non-linearities of the curvature perturbation. A formalism particularly useful when dealing with non-linearities is the so-called 
$\delta N$ formalism \cite{deltan}, where  the   scalar field fluctuations are quantized on the flat slices and  ${\cal R}=-\delta N$, being $N$ the number of e-folds.
The formalism is based on the  assumption that on super-Hubble scales, each spatial point of the universe has an independent evolution and the latter is well approximated by the evolution of an unperturbed universe.

Let us suppose that during the entire non-attractor phase the inflaton velocity decays exponentially. If so 
\be
\label{deltann}
{\cal N}(\phi,\Pi)=-\frac{1}{3}\ln\left[\frac{\Pi}{\Pi+3H(\phi-\phi_{\star})}\right]=-\frac{1}{3}\ln\frac{\Pi}{\Pi_\star},
\ee
where $\phi_\star$ is again the  value of the field at the end of the non-attractor phase.  Notice  that  we have retained the dependence on
$\Pi$ since slow-roll is badly violated. In  the relation (\ref{deltann}) we have followed the notation of Ref. \cite{sasaki} and defined ${\cal N}=0$ to be the end of the
attractor phase, so that ${\cal N}<0$ and
\begin{equation}
\phi({\cal N})=\phi_\star+\frac{\Pi_\star}{3}\left(1-e^{-3{\cal N}}\right)\,\,\,\,{\rm and}\,\,\,\, \Pi({\cal N})=\Pi_\star e^{-3{\cal N}}.
\end{equation}
We therefore find that
\be
\label{rt}
{\cal R}=-\delta{\cal N}=-{\cal N}+\overline{{\cal N}}=-\frac{1}{3}\ln\left(1+\frac{\delta\Pi_\star}{\overline{\Pi}_\star}\right),
\ee
where the overlines indicate the corresponding background values. One can safely neglect the perturbation $\delta\Pi$ as it decays exponentially fast. On the other
hand, by using the relation
%
\be
\Pi_\star=3\left[\phi({\cal N})-\phi_\star\right]+\Pi({\cal N}),
\ee
we see that up to irrelevant constants,
\be
\label{pu}
{\cal R}=-\frac{1}{3}\ln\left(1+3\frac{\delta\phi}{\overline{\Pi}_\star}\right),\,\,\,\,\delta\phi<-\frac{\Pi_\star}{3}.
\ee
The crucial point now is that the dynamics of $\delta\phi$ is the one of a massless perturbation in de Sitter and to a very good approximation its behaviour is Gaussian.
The non-Gaussianity in the curvature perturbation arises because of the non-linear mapping between $\delta\phi$ and ${\cal R}$\footnote{A few comments. The non-Gaussianity during the non-attractor phase is not washed out by the subsequent transition to a slow-roll phase. This is because
such a transition is sudden \cite{sasaki} as the velocity during the non-attractor phase must be much smaller than the one during the subsequent slow-roll phase to generate PBHs.
The non-Gaussianity we are dealing with here is not the  non-Gaussianity
in the squeezed configuration which peaks when   one of the wavelengths is  much larger than the other two. This non-Gaussianity is not observable by a local observer
testing a region much smaller than the long wavelength \cite{ngsqueezed}. We are instead referring to  that   non-Gaussianity which arises at the same small wavelengths  where  the density perturbations are sizeable. In the limit of a spiked power spectrum centered around a given momentum $k_{\rm pk}$, the non-Gaussianity will
be peaked at equilateral configurations.}.

The fact that $P(\delta\phi)$ is  Gaussian considerably  simplifies the computation:  the primordial mass fraction $\beta_{\rm prim}(M)$ of the universe occupied by PBHs  at formation time is dictated by probability conservation, 
\be
P({\cal R})=\left|\frac{\dd\delta\phi}{\dd{\cal R}}\right|P[\delta\phi({\cal R})], 
\ee
or
\begin{eqnarray}
P(\mathcal{R})=\frac{|\overline{\Pi}_\star|}{\sqrt{2\pi}\,\sigma_{\delta\phi}}
\exp\left[-3 \mathcal{R}-\frac{\overline{\Pi}_\star^2}{18\sigma_{\delta\phi}^2}\left(1-e^{-3\mathcal{R}}\right)^2\right],  \label{pp}
\end{eqnarray}
where we have written the Gaussian distribution of $\delta\phi$ as\footnote{Sometimes the Gaussian probability is multiplied by a factor of 2 to account for the fact that one deals with a first time-passage  problem \cite{mr}. We do not put it here as there is no general consensus of this factor. Quantitatively, it does not make a big difference though.}
\be
P(\delta\phi)=\frac{1}{\sqrt{2\pi}\,\sigma_{\delta\phi}}e^{-(\delta\phi)^2/2\sigma_{\delta\phi}^2}, 
\ee
with
\be
\sigma_{\delta\phi}^2=\int_{k}\dd\ln p\,{\cal P}_{\delta\phi}(p).
\ee
For $3\mathcal{R}_c\lsim 1$, we obtain
\begin{eqnarray}
P(\mathcal{R})\approx \frac{\overline{\Pi}_\star}{\sqrt{2\pi}\,\sigma_{\delta\phi}} e^{-\overline{\Pi}_\star^2\mathcal{R}^2/2\sigma_{\delta \phi}^2},
\end{eqnarray}
i.e., a Gaussian with variance 
\begin{eqnarray}
\sigma_{\mathcal{R}}^2=
\frac{\sigma_{\delta \phi}^2}{\overline{\Pi}_\star^2}.
\end{eqnarray}
On the other hand, assuming now $3{\cal R}_c\gsim 1$, we obtain
\begin{eqnarray}
\beta_{\rm prim} (M)&=&\int_{{\cal R}_{c}} {\rm d}{\cal R}\,P({\cal R})\simeq \frac{1}{2}{\rm erf}\left(\frac{\overline{\Pi}_\star}{3 \sqrt{2}\sigma_{\delta \phi}}\right)-\frac{1}{2}{\rm erf}\left(\frac{\overline{\Pi}_\star(1-e^{-3 \mathcal{R}_c})}{3\sqrt{2}\sigma_{\delta \phi}}\right)\nonumber\\
&\simeq &  -\frac{\Pi_\star e^{-3{\cal R}_c}}{3\sqrt{2\pi}\,\sigma_{\delta\phi}}e^{-\Pi^2_\star/18\sigma_{\delta\phi}^2},
\end{eqnarray}
to be confronted to the Gaussian result (\ref{ggg}). The probability is clearly non- Gaussian. We can estimate $\sigma_{\delta\phi}$ as well to be  of the order of $(H/2\pi)\Delta N$.  We obtain
\be
\beta_{\rm prim} (M)\simeq \frac{ e^{-3{\cal R}_c}}{3{\cal R}_\star\Delta N}e^{-1/(3\sqrt{2}{\cal R}_\star\Delta N)^2}.
\ee
To obtain the same primordial mass fraction, non-Gaussianity seems to require a smaller ${\cal R}_\star$. We write ``seems" as the curvature perturbation is not the best
variable to study the PBH mass function. As written in the main text, the density contrast $\Delta(\vec x)=(4/9a^2H^2)\nabla^2\zeta(\vec x)$ (during radiation) is the good  variable  \cite{bb}. This however  will make things more difficult to analyse 
because of the complication arising from taking the laplacian of the expression (\ref{rt}).
One possible, but not entirely satisfactory, way out might 
to evaluate  the density contrast at Hubble re-entry, i.e. setting $k=aH$. In such a case,  one could relate the density contrast to the curvature perturbation through the relation  $\Delta(\vec x)=(4/9){\cal R}(\vec x)$.

\section*{Appendix D: Comment on 1807.09057}
\setcounter{equation}{0}
\renewcommand{\theequation}{D.\arabic{equation}}
\noindent
After the publication  of this paper,  Ref.\cite{g} appeared with some comments about our findings. Here we  respond to them. This Appendix
can be considered as an independent part of this work and therefore some concepts of the main text might be repeated. 

First of all, to the best of our understanding, Ref. \cite{g}  just contains   the  demonstration that one can compute the curvature perturbation   in the non-attractor phase
using the stochastic approach instead of adopting the  standard computation in curved spacetime quantum field theory.  This result differs from that  in Ref. \cite{bellidodiffusion} and this is reassuring, as the standard linear computation has been our starting point\footnote{However, we stress that one can exactly map
by duality the non-attractor phase into a slow-roll phase in the limit of a plateau in the potential (see Ref.  \cite{wands} and Appendix A). This makes the  conclusions of Ref. \cite{g} suspicious, as they  claim that the non-attractor phase and the slow-roll phase behave differently at leading order in the slow-roll parameters.}. In our paper, however, we have not used the stochastic
approach to compute the perturbations, but to  investigate the role  of quantum diffusion on the observables. We did not {\it assume} the validity of the stochastic approach 
to compute the perturbations, the latter being derived by the standard field theory techniques (as in the large majority of the literature on the non-attractor phase).

The stochastic approach to study the cosmological perturbations focus on the behaviour of the {\it perturbations} on large scales under the action of the short modes which are integrated out from the action. These long mode perturbations are then treated classically under the action of a stochastic noise and give rise to a given power spectrum. In our approach, we do not focus  on the perturbations, but on the effect of the noise onto the {\it background} observables. 

Our results have therefore  little to do with those in Ref. \cite{g}. In fact, the importance of diffusion in the determination of the primordial PBH abundance has been already discussed in  Ref. \cite{ng6} where it was also shown to be crucial  (their analysis is restricted to slow-roll. However in the limit of extreme flatness of the potential during the non-attractor phase one can apply the duality discussed in Ref. \cite{wands} and in our Appendix A to map the problem into a slow-roll one). 

 Nevertheless, let us provide some comments about the criticisms raised in Ref. \cite{g}. This will also allow us to discuss  some  clarifications/considerations.

The  point raised in Ref. \cite{g} is that the power spectrum  is not a stochastic quantity and therefore may not be
used to calculate the impact of quantum diffusion onto the PBH abundance. However,   the smoothed power spectrum  (the one which enters in the calculation of the
abundance of PBHs) is a stochastic
quantity once one specifies the scale at which the average is operated and in the presence of long-mode perturbations with wavelengths larger than the size of the region where the
averaged is performed. This is nicely explained, for instance, in Ref. \cite{box}. Let us consider two counter-examples to the statement of  Ref. \cite{g}. First, the computation of the celebrated Maldacena consistency relation relating the power spectra of the curvature perturbation to the bispectrum in the squeezed limit. It is well-known that
such a result may be obtained simply taking the power spectrum, computed on a small box, and average it over a bigger volume containing the long-mode perturbation. The very simple result that the power spectrum correlates with the long mode shows that it is not a function, but a stochastic quantity. Similarly, in order to compute the
local halo bias in the presence of primordial non-Gaussianity one exploits the fact that the variance of the density contrast is  stochastic quantity in the presence of long-mode
perturbations.

As we explain in the main text, one way of producing PBHs in the early universe is to generate an enhancement of the power spectrum of the curvature perturbation during inflation, more specifically during the non-attractor. 
These large perturbations re-enter the horizon during the radiation era and may collapse to form PBHs on comoving scales that left the Hubble radius about $(20-30)$ e-folds before the end of inflation.

At the end of the  non-attractor phase  the curvature perturbation (in the flat gauge) is

\be
\label{a}
{\cal R}=-H\frac{\delta\phi}{\dot\phi_\star}
\ee
The calculation of the curvature perturbation till the end of  the non-attractor can be performed by using the $\delta N$ formalism \cite{sasaki}. 
Let us suppose that during the entire non-attractor phase the inflaton velocity decays exponentially so that

\begin{equation}
\label{b}
\phi({\cal N})=\phi_\star+\frac{{\dot\phi}_\star}{3H}\left(1-e^{-3{\cal N}}\right)\,\,\,\,{\rm and}\,\,\,\, {\dot\phi}({\cal N})={\dot\phi}_\star e^{-3{\cal N}}.
\end{equation} 
Here $\phi_\star$ and $\dot{\phi}_\star$ are  the   values of the inflaton field and its velocity at the end of the non-attractor phase, respectively  and we have set  ${\cal N}=0$ to be the end of the
attractor phase, so that ${\cal N}<0$. Then

\be
\label{deltann}
{\cal N}(\phi,{\dot\phi})=
-\frac{1}{3}\ln\frac{{\dot\phi}}{{\dot\phi}_\star}.
\ee
 On the other
hand, by using the relation (\ref{b}) and expanding at first-order one finds the expression (\ref{a}).
%
Now, in this paper we have followed the same logic which has been neatly explained   in Ref. \cite{s2}. The   $\delta {\cal N}$ method
 consists  of  three steps \cite{s2}:

\begin{itemize}
\item
First of all, to find  an inflationary trajectory for any point in the $(\phi,\dot\phi)$ space and  to calculate the number of e-folds ${\cal N}(\phi,\dot\phi)$ for this trajectory. 

\item  The position of the point $(\phi,\dot\phi)$ has to be perturbed  by adding to it inflationary jumps.  This provides  the perturbation of the number of e-folds $\delta {\cal N}$, which is directly related to the density perturbations. 

\item
The third step and most relevant for us (usually not performed in slow-roll single-field models) comes from the fact  that the resulting density perturbation for a given ${\cal N}$ (i.e. for a given wavelength) will depend on the place  $(\phi,\dot\phi)$ the  trajectory come from. Thus the remaining step is to evaluate the {\it probability} that for a given number of e-folds ${\cal N}$ till the end of the non-attractor phase (usually till the end of inflation in slow-roll models)  the field was at any particular point $(\phi,\dot\phi)$. This is because for an observer restricted to
her/his own Hubble radius during inflation, the classical value of the field is not given only by the zero mode, but also by the sum of the modes with wavelength larger than
the Hubble length. This was the essence of the results of Ref. \cite{ng6} which indeed found that the PBH abundance is  different when quantum diffusion is present.
The necessity of such third sanity-check step was  also stressed in Ref. \cite{box} (even though with no reference to PBHs).

As pointed out also in Ref. \cite{s2}, this third step  can be performed by using the stochastic approach   which tells what is the probability to find a given value of the inflaton field and its velocity at a  given point as a function of time.
\end{itemize}
In the main text we have pointed out that during the non-attractor  phase the role of quantum diffusion on the coarse-grained field $\dot\phi$  may become relevant and changes
the value measured by observers restricted to their own Hubble patches.  In other words, we have pointed out that the last and third sanity-check step described above is necessary. Usually  in  slow-roll single-field models  probabilities
are basically Gaussian   functions peaked around the classical values and tiny widths and one neglects the third step (even though for the problem of the PBH abundance is
indeed necessary \cite{ng6}).

In  order to compute the
variance of the sizeable curvature perturbation upon horizon re-entry, which will eventually give rise to PBHs by collapse,  one usually considers the
 classical evolution of the homogeneous fields $\phi$ and $\dot\phi$ and the effect of perturbations about the classical trajectories on a given scale.
However, in the extreme case in which diffusion overcomes the classicality, one may not  estimate  the curvature perturbation in terms of the classical trajectories \cite{s2}.
Luckily, in the case at hand, quantum diffusion never becomes more relevant  than  its classical evolution. Nevertheless,  even tiny differences may have an impact on the final abundance of the PBHs as it is exponentially sensitive to the  variance deduced from power spectrum 
of ${\cal R}$.

In order to  calculate the probability distribution for the field $\dot\phi$ we have used  the stochastic approach  which amounts to 
assuming an  average quantum diffusion per Hubble volume per Hubble time of the order of $H/2\pi$. The velocity of the inflaton field becomes also
a stochastic variable and the corresponding  variance  characterises the dispersion of the classical trajectories due to quantum fluctuations.

Taking for simplicity the case of constant potential during the non-attractor phase, the stochastic equations are (we report them here)

\begin{eqnarray}
H\frac{\partial}{\partial {\cal N}}\langle (\Delta\phi)^2\rangle&=&-2\langle\Delta\phi \Delta\dot\phi\rangle,\nonumber\\
H\frac{\partial}{\partial {\cal N}}\langle \Delta\phi \Delta\dot\phi\rangle&=&-\langle (\Delta\dot\phi)^2\rangle
+3\langle\Delta\phi \Delta\dot\phi\rangle,\nonumber\\
\frac{\partial}{\partial {\cal N}}\langle (\Delta\dot\phi)^2\rangle&=&6\langle (\Delta\dot\phi)^2\rangle
-H^2 D,
\end{eqnarray}
where $D=(3 H/2\pi)^2$ and we have indicated with $\Delta\phi=\phi-\phi({\cal N})$ and $\Delta\dot\phi=\dot\phi-\dot\phi({\cal N})$.
This set of equations shows that, even if the inflaton field and its velocity are taken to be homogeneous till one e-fold before the end of the
non-attractor phase, it is unavoidable that at the end of it the inflaton field receives kicks of the order of

\be
\Delta\phi \simeq \frac{H}{2\pi}
\ee
and the velocity of the order of

\be
\label{c}
\Delta\dot\phi\simeq \frac{\sqrt{3} H^2}{2\sqrt{2}\pi}.
\ee
In slow-roll one does not worry about these kicks, as the classical value of the velocity is given by $\dot{\phi}^2=2\epsilon H^2 M_p^2\gg (\Delta\dot\phi)^2\simeq H^4$, where $\epsilon\sim 10^{-2}$ is a slow-roll parameter
and $M_p$ the reduced Planck mass. This effect is totally negligible.  However, at the end of the non-attractor phase $\epsilon$ is much smaller, $\epsilon\sim 10^{-8}$ . The variance of the inflaton velocity is not negligible when  recalling  
that tiny changes of the curvature perturbations are exponentially inflated when computing the PBH abundance. 

Notice that the variance of the inflaton velocity reaches the value  (\ref{c}) after one e-fold or so, and remains the same on all coarse-grained
lengths. In particular, assuming that the peak of the perturbation is reached, say, 20 e-folds before the end of inflation, our current universe will contain, at the time
of formation of the PBHs, about $\exp(40\cdot 3)=\exp(120)$ Hubble volumes. In any of them the velocity has tiny differences due to the variance (\ref{c}).
Since the probability to form a PBH in any of each patches is Eq. (1.1), 
having a not fully  fixed ${\cal R}$ leads to different values of $\sigma_{{\cal R}}^2$ and therefore $\beta$ in all the patches. One point to 
stress is that the kicks (\ref{c}) are kicks of the short modes leaving the Hubble radius each Hubble time and what they do is to change the infrared long modes
of the inflaton velocity whose cumulative effect is  measured by the local observer as  the background  inflaton velocity. 

Thus, while the final word is certainly given by the calculation of the exact probability of the comoving curvature perturbation and the  corresponding KM-like equation (see for example Ref.  \cite{rs} for the slow-roll case),  in order to avoid any analytical approximation, we have numerically constructed different realisations of the comoving curvature perturbation by solving the corresponding equation of motion (A.1) for any given wavenumber,  one for each random trajectory identified by the corresponding coarse-grained background values. We have then deduced the mean and the variance 
of the abundance of the PBHs. These procedures are equivalent.

As a final note, let us stress that the fact that the density contrast barrier depends on the shape of the power spectrum (which in turn determines the shape of the perturbation in real space collapsing to a PBH) implies that a change in the comoving curvature perturbation in each Hubble patch at horizon re-entry 
due to quantum diffusion will determine a different barrier. This as well is expected to have an impact on the final PBH abundance distribution.

\end{document}